\documentclass[12pt]{article}

\usepackage[a4paper, left=2.5cm, right=2.5cm, bottom=4.5cm, heightrounded]{geometry}

\usepackage[UKenglish]{babel}

\usepackage{cite}
\usepackage{comment}            
\usepackage{scalerel}
\usepackage{shuffle}
\usepackage{latexsym,epsfig}
\usepackage[hidelinks]{hyperref}
\usepackage{dsfont}
\usepackage{nccmath}            
\usepackage{nicefrac}
\usepackage{tcolorbox}
\usepackage{multirow}           
\usepackage{longtable}          

\usepackage{amsmath,amsfonts,amssymb,mathtools}
\usepackage{extarrows}
\usepackage{cancel, slashed}

\usepackage{upgreek}            
\usepackage[bbgreekl]{mathbbol} 

\DeclareSymbolFontAlphabet{\mathbbg}{bbold} 
\DeclareSymbolFontAlphabet{\mathbb}{AMSb}   

\usepackage{tikz, tikz-cd}

\usepackage{xcolor}

\numberwithin{equation}{section}
\allowdisplaybreaks

\def\bpm{\begin{pmatrix}}
\def\epm{\end{pmatrix}}
\newcommand{\defeq}{{\,\coloneq\,}}         
\newcommand{\eqdef}{{\,\eqcolon\,}}         

\newcommand{\mycomment}[1]{}
    \def\cA{{\cal A}}
    \def\cB{{\cal B}}
    
    \def\cD{{\cal D}}

    \def\cJ{{\cal J}}

    \def\cM{{\cal M}}

    \def\cT{{\cal T}}

    \def\bbC{\mathbb{C}}

    \def\bbP{\mathbb{P}}

    \def\del{\partial}

\begin{document}

\begin{flushright}
    HU-EP-25/40-RTG
\end{flushright}

\vskip 13mm

\begin{center}
    
    {\large\textbf{Supergravity realisations of $\lambda$-models}}
    
    \vskip 10mm

    Giuseppe Casale$^1$, Georgios Itsios$^2$

    \vskip 10mm

    {\small\textit{Institut f\"ur Physik, Humboldt-Universit\"at zu Berlin}} \\
    {\small\textit{Zum Großen Windkanal 2, D-12489 Berlin, Germany}}

    \vskip 25mm

    {\bf{Abstract}}

\end{center}

\begin{quote}
    We construct solutions of type-II supergravity based on multiple copies and/or mixings of $\lambda$-deformed coset CFTs on $\nicefrac{\mathrm{SO}(n+1)_k}{\mathrm{SO}(n)_k}$, with $n = 2, 3, 4$. The resulting ten-dimensional geometries contain undeformed $\mathrm{AdS}$ factors, thereby allowing a connection between $\lambda$-deformations and the AdS/CFT correspondence. Imposing reality conditions on the solutions further constrains the deformation parameter. In some cases these bounds exclude the undeformed ($\lambda = 0$) or non-Abelian T-dual ($\lambda \to 1$) limits. This work extends the results of [\href{https://arxiv.org/abs/1911.12371}{1911.12371}] and [\href{https://arxiv.org/abs/2411.11086}{2411.11086}].
\end{quote}

\vfill

\noindent
{\small
$^1$ giuseppe.casale@physik.hu-berlin.de \\
$^2$ georgios.itsios@physik.hu-berlin.de
}

\thispagestyle{empty}

\newpage

\clearpage
\pagenumbering{arabic}

\tableofcontents

\setcounter{footnote}{0}

\section{Introduction}
\label{Sec:Intro}

While integrability-preserving deformations of non-linear $\sigma$-models are interesting in their own right, their interplay with ten-dimensional supergravity and their applications in holography provide additional motivation for studying them. An example that stands out within this class of models is the family of $\lambda$-deformations, introduced in \cite{Sfetsos:2013wia}. In their original formulation, they describe a two-dimensional field theory that interpolates between an exact conformal field theory (CFT) -- a Wess--Zumino--Witten (WZW) model \cite{Witten:1983ar} on a semisimple group $G$ with current algebra symmetry at level $k$ -- and the non-Abelian T-dual (NATD) of a principal chiral model (PCM) on $G$. This construction has been generalised in various directions. Deformations based on cosets with a Lagrangian description in terms of gauged WZW models, as well as those associated with symmetric or semi-symmetric spaces, were developed in \cite{Sfetsos:2013wia,Hollowood:2014rla,Hollowood:2014qma}. A variation of the $\lambda$-model that incorporates spectator degrees of freedom was derived in \cite{Borsato:2024alk}. Further generalisations include \cite{Georgiou:2016zyo,Georgiou:2017jfi,Georgiou:2018hpd,Georgiou:2018gpe,Driezen:2019ykp,Sfetsos:2017sep}. Their relation to another class of integrable $\sigma$-models, known as $\eta$-deformations \cite{Klimcik:2002zj,Klimcik:2008eq}, which are based on a PCM for some group, was established in \cite{Vicedo:2015pna,Hoare:2015gda,Sfetsos:2015nya}.

The connection between ten-dimensional supergravity and integrable deformations of $\sigma$-models arises by promoting the corresponding $\sigma$-model background fields to full solutions of type-II supergravity. This has been achieved for $\lambda$- and $\eta$-deformed (super)cosets in various works \cite{Demulder:2015lva,Hoare:2015gda,Borsato:2016zcf,Itsios:2019izt,Itsios:2024gqr,Chervonyi:2016ajp,Borsato:2016ose,Hoare:2015wia,Hoare:2018ngg,Lunin:2014tsa,Hoare:2022asa}. Deformations of the near-horizon limit of $\mathrm{NS1}$ and $\mathrm{NS5}$ brane intersections have been constructed recently in \cite{Itsios:2023kma,Itsios:2023uae}. These backgrounds accommodate the $\lambda$-models on $\mathrm{SL}(2,\mathbb{R})$ and $\mathrm{SU}(2)$ within a ten-dimensional setting.

The main objective of the present work is to extend the results of \cite{Itsios:2019izt,Itsios:2024gqr} and construct type-II supergravity backgrounds incorporating multiple copies and/or mixings of $\lambda$-deformed coset CFTs on $\nicefrac{\mathrm{SO}(n+1)_k}{\mathrm{SO}(n)_k}$, with $n = 2 , 3 , 4$. Our method relies on proposing educated ans\"atze for the corresponding Ramond--Ramond (RR) fields. The advantage of this approach is that the problem of finding solutions to the supergravity equations of motion reduces to solving algebraic systems of constant parameters, rather than non-linear PDEs. Motivated by holography, we focus on geometries that exhibit undeformed $\mathrm{AdS}$ factors. In particular, our goal is to construct solutions with geometry of the direct-product form
\begin{equation}
\label{GeometriesInGeneral}
 \cM_{10 - d} \times {\rm CS}^{n_1}_{\lambda} \times \ldots \times {\rm CS}^{n_k}_{\lambda} \, , 
 \qquad d = n_1 + \ldots + n_k \, .    
\end{equation}
We will focus on cases in which the space $\cM_{10 - d}$ satisfies either of the following properties
\begin{enumerate}
    \item It is a $(10 - d)$-dimensional AdS space.
    \item It is a direct product of the AdS space with Einstein manifolds, such as tori, spheres, hyperbolic spaces, and complex projective spaces.
\end{enumerate}
\noindent Our findings are summarised in Table~\ref{Tab:Geometry&Sugra}.

\begin{longtable}{|c|c|c|c|}
    \hline
        \textbf{Geometry} & \textbf{Range of $\lambda$} & \textbf{Supergravity} & \textbf{Example} \\
    \hline\hline
        $\rm AdS_6 \times CS^2_\lambda \times CS^2_\lambda$                         & \multirow{5}{*}{$[0,\frac12(3-\sqrt5)]$} & \multirow{5}{*}{IIA}  & \multirow{5}{*}{Sec. \ref{Subsubsec:IIA_M6CS2CS2} Ex. 1} \\*
        $\rm AdS_4 \times H_2 \times CS^2_\lambda \times CS^2_\lambda$              &  &  &  \\*
        $\rm AdS_3 \times H_3 \times CS^2_\lambda \times CS^2_\lambda$              &  &  &  \\*
        $\rm AdS_2 \times H_4 \times CS^2_\lambda \times CS^2_\lambda$              &  &  &  \\*
        $\rm AdS_2 \times H_2 \times H'_2 \times CS^2_\lambda \times CS^2_\lambda$  &  &  &  \\
    \hline
        $\rm AdS_4 \times S^2 \times CS^2_\lambda \times CS^2_\lambda$              & $[0,1)$ & \multirow{2}{*}{IIA}  & \multirow{2}{*}{Sec. \ref{Subsubsec:IIA_M6CS2CS2} Ex. 2} \\*
        $\rm AdS_4 \times H_2 \times CS^2_\lambda \times CS^2_\lambda$              & $[0,2-\sqrt3]$ &  & \\
    \hline
        $\rm AdS_2 \times T^4 \times CS^2_\lambda \times CS^2_\lambda$              & \multirow{3}{*}{[0,1)}  & \multirow{3}{*}{IIA}  & \multirow{3}{*}{Sec. \ref{Subsubsec:IIA_M6CS2CS2} Ex. 3} \\*
        $\rm AdS_2 \times \mathbb{CP}^2 \times CS^2_\lambda \times CS^2_\lambda$              &  &  &  \\*
        $\rm AdS_2 \times S^4 \times CS^2_\lambda \times CS^2_\lambda$              &  &  &  \\*
        $\rm AdS_2 \times S^2 \times S'^2 \times CS^2_\lambda \times CS^2_\lambda$  &  &  &  \\
    \hline
        $\rm AdS_2 \times \bbC\bbP^2 \times CS^2_\lambda \times CS^2_\lambda$       & $[0,1)$ & IIA  & Sec. \ref{Subsubsec:IIA_M6CS2CS2} Ex. 4 \\
    \hline
        $\rm AdS_2 \times T^2 \times CS^2_\lambda \times CS^2_\lambda \times CS^2_\lambda$ & $[0,1)$ & IIA & Sec. \ref{Subsubsec:IIA_M4CS2CS2CS2} \\
    \hline
        $\rm AdS_2 \times S^2 \times CS^2_\lambda \times CS^2_\lambda \times CS^2_\lambda$ & \multirow{4}{*}{$[0,1)$} & \multirow{4}{*}{IIB} & \multirow{4}{*}{Sec. \ref{Subsubsec:IIB_M4CS2CS2CS2} Ex. 1 -- 4} \\*
        $\rm AdS_2 \times T^2 \times CS^2_\lambda \times CS^2_\lambda \times CS^2_\lambda$ &  &  &  \\*
        $\rm AdS_2 \times H_2 \times CS^2_\lambda \times CS^2_\lambda \times CS^2_\lambda$ &  &  &  \\*
        $\rm AdS_4 \times CS^2_\lambda \times CS^2_\lambda \times CS^2_\lambda$ &  &  &  \\
    \hline
        $\rm AdS_2 \times CS^2_\lambda \times CS^2_\lambda \times CS^2_\lambda \times CS^2_\lambda$ & $[0,1)$ & IIA & Sec. \ref{Subsubsec:IIA_M2CS2CS2CS2CS2} \\
    \hline\hline
        $\rm AdS_4 \times CS^3_\lambda \times CS^3_\lambda$ & \multirow{3}{*}{$[0,1)$} & \multirow{3}{*}{IIA} & \multirow{3}{*}{Sec. \ref{Subsubsec:IIA_M4CS3CS3} Ex. 1 -- 2} \\*
        $\rm AdS_2 \times H_2 \times CS^3_\lambda \times CS^3_\lambda$ &  &  &  \\*
        $\rm AdS_3 \times S^1 \times CS^3_\lambda \times CS^3_\lambda$ &  &  &  \\
    \hline
        $\rm AdS_2 \times CS^4_\lambda \times CS^4_\lambda$ & $[0,1)$ & IIA & Sec. \ref{Subsubsec:IIA_M2CS4CS4} \\
    \hline\hline
        $\rm AdS_2 \times T^2 \times CS^2_\lambda \times CS^4_\lambda$ & $[0,1)$ & IIA & Sec. \ref{Subsubsec:IIA_M4CS2CS4} \\
    \hline
        $\rm AdS_2 \times S^2 \times CS^2_\lambda \times CS^4_\lambda$ & $[0,1)$ & \multirow{4}{*}{IIB} & \multirow{4}{*}{Sec. \ref{Subsubsec:IIB_M4CS2CS4} Ex. 1 -- 4} \\*
        $\rm AdS_2 \times T^2 \times CS^2_\lambda \times CS^4_\lambda$ & $[0,1)$ &  &  \\*
        $\rm AdS_2 \times H_2 \times CS^2_\lambda \times CS^4_\lambda$ & $[0,1)$ &  &  \\
    \hline
        $\rm AdS_2 \times CS^2_\lambda \times CS^2_\lambda \times CS^4_\lambda$ & $[0,1)$ & IIA & Sec. \ref{Subsubsec:IIA_M2CS2CS2CS4} \\
    \hline\hline
        $\rm AdS_2 \times S^2 \times \bbC\bbP^2 \times CS^2_\lambda$ & $(2-\sqrt3,1)$ & \multirow{4}{*}{IIB} & \multirow{4}{*}{Sec. \ref{Subsubsec:IIB_M4CP2CS2} Ex. 1 -- 4} \\*
        $\rm AdS_2 \times T^2 \times \bbC\bbP^2 \times CS^2_\lambda$ & $(2-\sqrt3,1)$ &  &  \\*
        $\rm AdS_2 \times H_2 \times \bbC\bbP^2 \times CS^2_\lambda$ & $(2-\sqrt3,1)$ &  &  \\*
        $\rm AdS_4 \times \bbC\bbP^2 \times CS^2_\lambda$ & $[0,1)$ &  &  \\
    \hline
        $\rm AdS_2 \times \bbC\bbP^2 \times CS^4_\lambda$ & $[0,1)$ & IIA & Sec. \ref{Subsubsec:IIA_M2CP2CS4} \\
    \hline
    \caption{Type-II backgrounds and their locations in the main text and appendix.}
    \label{Tab:Geometry&Sugra}
\end{longtable}

The rest of the paper is organised as follows. In Sec.~\ref{Sec:Review_lambda_def} we review the $\sigma$-model field content for the $\lambda$-deformed cosets on $\nicefrac{\mathrm{SO}(n+1)_k}{\mathrm{SO}(n)_k}$, with $n = 2, 3, 4$. Sec.~\ref{Sec:Sugra_Sols} contains the details of the construction of the type-II backgrounds. Conclusions and future directions are presented in Sec.~\ref{Sec:Conclusions}. 

We have also included two appendices. Appendix~\ref{Sec:Appendix_Eoms} reviews the equations of motion for type-II supergravity. Appendix~\ref{Sec:Appendix_Single_CS2s} discusses solutions built from a single copy of the $\lambda$-models on $\nicefrac{\mathrm{SO}(3)_k}{\mathrm{SO}(2)_k}$ and $\nicefrac{\mathrm{SO}(5)_k}{\mathrm{SO}(4)_k}$, whose geometry also exhibits a $\mathbb{CP}^2$ space.

\section{Review of $\lambda$-deformed coset CFTs}
\label{Sec:Review_lambda_def}
In this section we review the field content of $\lambda$-deformed models \cite{Sfetsos:2013wia}, built on cosets that involve orthogonal groups \cite{Demulder:2015lva}, which will be denoted as
\begin{align}
    {\rm CS}^n_\lambda \defeq \frac{{\rm SO}(n+1)_k}{{\rm SO}(n)_k} \;.
\end{align}
Here $k$ stands for the WZW level, and $\lambda$ is the deformation parameter which takes values in the interval $[0,1)$. In particular, we will focus on the cases of $n=2,3,4$, with target space always expressed by means of a Euclidean signature.

\subsection{The ${\rm CS}^2_\lambda$ model}
\label{Subsec:CS^2_Model}
Let us start by introducing the simplest case, namely the $\nicefrac{{\rm SO}(3)_k}{{\rm SO}(2)_k}$ model, which corresponds to the $\sigma$-model with a 2-dimensional target space with coordinates $x, \omega$ and a vanishing antisymmetric field.
For convenience we use the frame
\begin{subequations}\label{Eq:frame2}\begin{align}
    {\frak e}^1 &= -2\,\lambda_- \left(\frac{dx}{\omega}-\frac{x\,d\omega}{\omega_+^2}\right) \\
    {\frak e}^2 &= -2\,\lambda_+ \left(\frac{x\,dx}{\sqrt{\cal D}\,\omega}+\frac{\sqrt{\cal D}\,d\omega}{\omega_+^2}\right)
\end{align}\end{subequations}
where the functions $\cal{D} , \omega_{\pm}$ and constants $\lambda_{\pm}$ are defined below
\begin{align}\label{Eq:functions2}
    {\cal D} \defeq 1 - x^2 \;, \quad
    \omega_\pm \defeq \sqrt{1\pm\omega^2} \;, \quad
    \lambda_\pm \defeq \sqrt{k\, \frac{1 \pm \lambda}{1 \mp \lambda}} \;.
\end{align}
In addition, the model includes the scalar
\begin{equation}\label{scalar2}
    \Phi = -\frac12 \log \left( \frac{2\omega^2}{\omega_+^2} \right) \;.
\end{equation}
The above geometry and scalar satisfy the following relations
\begin{subequations}\label{betaFuncs2}\begin{align}
    &\beta_\Phi = R + 4\nabla^2\Phi - 4(\del\Phi)^2 = \nu \label{betaScalar2}\\
    &\beta_{g_{ab}} = R_{ab} + 2\nabla_a\nabla_b\Phi = - \mu \, \eta_{ab} \label{betaMetric2}
\end{align}\end{subequations}
where for later convenience we defined the constants
\begin{equation}\label{defmunu}
 \mu \defeq \frac{1}{k}\frac{\lambda}{1-\lambda^2} \;, \quad \nu \defeq \frac{1}{k}\frac{1+\lambda^2}{1-\lambda^2} \; .
\end{equation}
Moreover, $\eta_{ab}={\rm diag}(-1,+1)$, and $R$ and $R_{ab}$ are the Ricci scalar and the Ricci tensor, respectively. The above expressions are computed in the frame \eqref{Eq:frame2}.
It is clear that in the absence of deformation (i.e. when $\lambda=0$) $\beta_{g_{ab}}=0$ and the model becomes conformal. The underlying $\sigma$-model is invariant under
\begin{equation}\label{symmetrylambdak}
    \lambda \mapsto \lambda^{-1} \; , \qquad k \mapsto -k \; .
\end{equation}
This symmetry is also manifest in the frame \eqref{Eq:frame2}, as well as in $\beta_\Phi$ and $\beta_{g_{ab}}$. Moreover, the following identities hold
\begin{equation}\label{conditionsCS2lambda}
    d(e^{-\Phi}{\frak e}^1) =  d(e^{-\Phi}{\frak e}^2) = 0
\end{equation}
and will prove useful for the construction of the ans\"atze for the RR fields.

\subsection{The ${\rm CS}^3_\lambda$ model}
\label{Subsec:CS^3_Model}
Moving to the 3-dimensional case $\nicefrac{{\rm SO}(4)_k}{{\rm SO}(3)_k}$, we can express the target space geometry with coordinates $x, y, \omega$ via the following new set of dreibein
\begin{subequations}\label{Eq:frame3}\begin{align}
    {\frak e}^1 &= \frac{2\,\lambda_-}{\sqrt{\cA}} \left(-\omega x\,dx + \frac{y}{\omega}dy + \frac{\cA}{\omega^2_+}d\omega \right) \\
    {\frak e}^2 &= \frac{2\,\lambda_+}{\sqrt{\cA\cD}} \left(\cD\omega\,dx - \frac{xy}{\omega}dy + \frac{\cA x}{\omega^2_-}d\omega \right) \\
    {\frak e}^3 &= \frac{2\,\lambda_+}{\sqrt{\cD}} \left(\frac{dy}{\omega} + \frac{y}{\omega^2_-}d\omega \right)
\end{align}\end{subequations}
where $\cal D$, $\omega_\pm$ and $\lambda_\pm$ are given by \eqref{Eq:functions2}, while
\begin{equation}\label{Eq:functions3}
    {\cal A} \defeq 1-x^2-y^2 \;.
\end{equation}
Again, the model has a vanishing antisymmetric field and it is equipped with a scalar, which in this case reads
\begin{equation}\label{scalarCS3}
    \Phi = -\frac12 \log \left(8\frac{{\cal A}\omega^2}{\omega_+^4}\right) \;.
\end{equation}
The corresponding $\beta_\Phi$ and $\beta_{g_{ab}}$ are now given by
\begin{subequations}\label{betaFuncs3}\begin{align}
    &\beta_\Phi = R + 4\nabla^2\Phi - 4(\del\Phi)^2 = 3\nu \;, \label{betaScalar3}\\
    &\beta_{g_{ab}} = R_{ab} + 2\nabla_a\nabla_b\Phi = - 2\mu \, \hat\eta_{ab} \;, \label{betaMetric3}
\end{align}\end{subequations}
where $\mu , \nu$ are given in \eqref{defmunu} and $\hat\eta_{ab} = {\rm diag}(-1,+1,+1)$. Both $\beta_\Phi$ and $\beta_{g_{ab}}$, as well as the frame \eqref{Eq:frame3} respect the symmetry \eqref{symmetrylambdak}. Like in the previous example, $\beta_{g_{ab}}$ becomes trivial when $\lambda = 0$ and therefore the model is conformal. Additionally, the following identities are true for the frame \eqref{Eq:frame3}
\begin{equation}\label{conditionsCS3lambda}
    d(e^{-\Phi}{\frak e}^1) = 0 \;, \quad
    d(e^{-\Phi}{\frak e}^1\wedge{\frak e}^3) = d(e^{-\Phi}{\frak e}^2\wedge{\frak e}^3) = 0 \;.
\end{equation}
The first and the third identities will be crucial for the construction of type-II supergravity solutions.

\subsection{The ${\rm CS}^4_\lambda$ model}
\label{Subsec:CS^4_Model}
We now move to the last case, which corresponds to the $\lambda$-deformation of the $\nicefrac{{\rm SO}(5)_k}{{\rm SO}(4)_k}$ WZW model. Here the target space with coordinates $x, y, z, \omega$ is spanned by the vierbein
\begin{subequations}\label{Eq:frame4}\begin{align}
    {\frak e}^1 &= -\frac{2\,\lambda_-}{\sqrt{\cA\cB\cD}} \left(\frac{\cB\cD}{\omega y}dx + x \left(\frac{\cD z^2}{\omega y^2} + \omega y^2\right)dy - \frac{\cA xz}{\omega y}dz - \frac{\cA\cB}{\omega_+^2 y}d\omega \right) \;, \\
    {\frak e}^2 &= \frac{2\,\lambda_+}{\sqrt{\cA\cB}} \left( \frac{\cB x}{\omega y}dx + \left(\frac{x^2z^2}{\omega y^2} - \omega y^2\right)dy + \frac{\cA z}{\omega y}dz + \frac{\cA\cB}{\omega_+^2 y}d\omega \right) \;, \\
    {\frak e}^3 &= \frac{2\,\lambda_-}{\sqrt{\cB\cD}} \left( \omega y\,dy + \frac{z}{\omega}dz + \frac{\cB}{\omega_+^2}d\omega \right) \;, \\
    {\frak e}^4 &=  -2\,\lambda_+ \left( \frac{dz}{\omega y} - \frac{z}{\omega_+^2 y}d\omega \right) \;,
\end{align}\end{subequations}
with $\cD$, $\omega_\pm$ and $\lambda_\pm$ defined as in \eqref{Eq:functions2}, $\cA$ defined as in \eqref{Eq:functions3} and
\begin{equation}
    \cB \defeq y^2 - z^2 \;.
\end{equation}
As it is typical for this family of models, the antisymmetric field is trivial. The scalar characterising the model is given by
\begin{equation}\label{scalarCS4}
    \Phi = - \frac12 \log \left(64\frac{\cA\cB\omega^4}{\omega_+^6} \right)
\end{equation}
and -- together with the metric -- fulfils the relations
\begin{subequations}\label{betaFuncs4}\begin{align}
    &\beta_\Phi = R + 4\nabla^2\Phi - 4(\del\Phi)^2 = 6 \nu \;, \label{betaScalar4}\\
    &\beta_{g_{ab}} = R_{ab} + 2\nabla_a\nabla_b\Phi = - 3 \mu \, \tilde\eta_{ab} \;, \label{betaMetric4}
\end{align}\end{subequations}
with $\mu, \nu$ given in \eqref{defmunu} and $\tilde\eta_{ab} = {\rm diag}(+1,-1,+1,-1)$. Equation \eqref{symmetrylambdak} still provides a symmetry of the $\beta$-functions above, and the frame \eqref{Eq:frame4}. Again, a vanishing $\lambda$ would reduce to the conformal model.
Like the previous two examples, the frame satisfies the following identities
\begin{equation}\label{conditionsCS4lambda}
    d(e^{-\Phi}{\frak e}^1\wedge{\frak e}^3) = d(e^{-\Phi}{\frak e}^2\wedge{\frak e}^4) = 0 \;, \quad
    d(e^{-\Phi}{\frak e}^1\wedge{\frak e}^3\wedge{\frak e}^4) = 0 \;.
\end{equation}
In the next sections we move to the explicit construction of supergravity solutions focusing on geometries containing undeformed $\rm AdS$ factors, and in particular of the form \eqref{GeometriesInGeneral}. We will see that the properties \eqref{betaFuncs2}, \eqref{conditionsCS2lambda}, \eqref{betaFuncs3}, \eqref{conditionsCS3lambda}, \eqref{betaFuncs4} and \eqref{conditionsCS4lambda} enable us to propose suitable ans\"atze for the supergravity fields.

\section{Supergravity solutions}
\label{Sec:Sugra_Sols}
In this section, we promote multiple copies and/or mixings of the ${\rm CS}^n_{\lambda}$ ($n = 2, 3, 4$) models into full solutions of the type-II supergravities. A complete account of our findings is summarised in table \ref{Tab:Geometry&Sugra} in the Introduction.

\subsection{Solutions involving only ${\rm CS}^2_\lambda$}
\label{Subsec:Only_CS2}

We begin by discussing the construction of type-II backgrounds from multiple copies of the two-dimensional model $\rm CS^2_\lambda$, introduced in Sec. \ref{Subsec:CS^2_Model}. The same ideas will be applied to the higher dimensional cases $\rm CS^3_\lambda$ and $\rm CS^4_\lambda$.


\refstepcounter{subsubsection}\subsubsection*{\thesubsubsection\quad Type-IIA on $\cM_6 \times {\rm CS}^2_\lambda \times {\rm CS}^2_\lambda$}
\label{Subsubsec:IIA_M6CS2CS2}

As a first example, we consider the case with two copies of ${\rm CS}^2_{\lambda}$. We will assume that the ten-dimensional geometry takes the direct product form $\cM_6 \times {\rm CS}^2_{\lambda} \times {\rm CS}^2_{\lambda}$, where the properties of the six-dimensional manifold $\cM_6$ will be determined later. Since the target spaces on ${\rm CS}^2_{\lambda}$ have Euclidean signature, we expect that the time-like direction sits in $\cM_6$ and the corresponding line-element in terms of the frame ($e^0 , \ldots , e^5$) reads
\begin{equation}
    ds^2_{\cM_6} = - (e^0)^2 + (e^1)^2 + \ldots + (e^5)^2 \;.
\end{equation}
The other four directions of the ten-dimensional geometry are identified as $e^6 \to \mathfrak{e}^1$, $e^7 \to \mathfrak{e}^2$, $e^8 \to \mathfrak{e}^1$ and $e^9 \to \mathfrak{e}^2$, with $\mathfrak{e}^1$ and $\mathfrak{e}^2$ given in \eqref{Eq:frame2}. It is clear that in order to distinguish the two ${\rm CS}^2_{\lambda}$ spaces, one has to adopt a different labelling for the coordinates $(x, \omega)$. In particular, we will use the coordinates $(x_1 , \omega_1)$ for the ${\rm CS}^2_{\lambda}$ space spanned by $(e^6 , e^7)$, and the coordinates $(x_2 , \omega_2)$ for the ${\rm CS}^2_{\lambda}$ space spanned by $(e^8 , e^9)$. A second assumption that we make is that we take the Neveu-Schwarz (NS) two-form to be trivial. As for the dilaton, we assume that it is given by the sum of two copies of the scalar \eqref{scalar2}
\begin{equation}
 \Phi = \Phi_1(\omega_1) + \Phi_2(\omega_2) = -\frac12 \log \left( \frac{2\omega_1^2}{\omega_{1+}^2} \right) - \frac12 \log \left( \frac{2\omega_2^2}{\omega_{2+}^2} \right) \;.
\end{equation}
The above choice for the dilaton -- namely, taking it to be the sum of the two scalars -- and the assumption of a vanishing NS two-form are motivated by the dilaton equation \eqref{dilatonEOM} and the property \eqref{betaScalar2}, in view of the fact that the ten-dimensional geometry is a direct product of subspaces. As we will see shortly, this allows us to express eq.~\eqref{dilatonEOM} as the sum of the beta-functions $\beta_{\Phi}$ associated with each of the $\lambda$-models appearing in the embedding, which are constants, together with the Ricci scalar of the external space $\cM_6$. Likewise, the Einstein equations \eqref{EinsteinEOMs} simplify significantly due to the property \eqref{betaMetric2}. We will also see that this choice of dilaton enables us to exploit the properties \eqref{conditionsCS2lambda} in order to construct suitable ansätze for the RR fields. Although one could in principle turn on an NS two-form, we refrain from doing so, as this would introduce additional complications in the dilaton equation \eqref{dilatonEOM}, the Einstein equations \eqref{EinsteinEOMs}, and the field equation \eqref{fieldeqnH}, where its field strength $H$ couples with the dilaton. The same reasoning will be employed in all examples throughout the text.

By combining the property \eqref{betaScalar2} with the equation of motion for the dilaton \eqref{dilatonEOM}, one can deduce the curvature on $\cM_6$
\begin{equation}\label{curvatureM6}
    R_{\cM_6} = - 2 \nu \;,
\end{equation}
which is constant and negative for all values of $\lambda$ in $[0 , 1)$.

We can learn more about the structure of $\cM_6$ if we introduce a specific ansatz for the RR fields. In this case we will consider
\begin{subequations}
\label{RRansatzEx311}
 \begin{align}
  F_2 & = 2 e^{-\Phi} \big( c_1 e^{68} + c_2 e^{69} + c_3 e^{78} + c_4 e^{79} \big) \;,
  \\
  F_4 & = 2 e^{-\Phi} e^{01} \wedge \big( c_5 e^{68} + c_6 e^{69} + c_7 e^{78} + c_8 e^{79} \big)
  \nonumber \\
  & + 2 e^{-\Phi} e^{23} \wedge \big( c_9 e^{68} + c_{10} e^{69} + c_{11} e^{78} + c_{12} e^{79} \big)
  \\
  & + 2 e^{-\Phi} e^{45} \wedge \big( c_{13} e^{68} + c_{14} e^{69} + c_{15} e^{78} + c_{16} e^{79} \big) \;, \nonumber
 \end{align}
\end{subequations}
where $c_i \, (i = 1 , \ldots , 16)$ are taken to be constants. We adopt the convention $e^{a_1\dots a_p} \defeq e^{a_1}\wedge\dots\wedge e^{a_p}$ to simplify our ans\"atze.

The rationale behind the above choice lies in the specific form of the dilaton, written as the sum of the two scalars $\Phi_1$ and $\Phi_2$, in combination with the properties \eqref{conditionsCS2lambda}. In particular, we exploit these properties in order to construct a \emph{closed} form supported entirely on the internal space, namely on the ${\rm CS}^2_{\lambda} \times {\rm CS}^2_{\lambda}$ sector of the geometry, whose Hodge dual (taken within the internal space) is also closed. In the present example, and in view of the properties \eqref{conditionsCS2lambda}, this form is a two-form given as a linear combination of $e^{-\Phi} e^{68}$, $e^{-\Phi} e^{69}$, $e^{-\Phi} e^{78}$, and $e^{-\Phi} e^{79}$, with constant relative coefficients\footnote{
In view of Eq.~\eqref{conditionsCS2lambda}, one can show that
\[
d \left( e^{-\Phi} e^{68} \right) 
= \cancelto{0}{d \left( e^{-\Phi_1} e^6 \right)} \wedge \left( e^{-\Phi_2} e^8 \right) 
- \left( e^{-\Phi_1} e^6 \right) \wedge \cancelto{0}{d \left( e^{-\Phi_2} e^8 \right)} 
= 0 \, .
\]
An analogous argument applies to $e^{-\Phi} e^{69}$, $e^{-\Phi} e^{78}$, and $e^{-\Phi} e^{79}$, as well as to their Hodge duals within the internal space.}.
In type-IIA supergravity, this two-form can serve as the RR flux $F_2$. A four-form flux $F_4$ may then be constructed by taking wedge products of this closed internal two-form with two-forms supported on the external space $\cM_6$. As we will show below, the type-IIA equations of motion for the RR fluxes impose constraints on the transverse space $\cM_6$. It is also important to note that the presence of the factor $e^{-\Phi}$ in the ansatz ensures that $\cT_{\mu\nu}$ in eq.~\eqref{EinsteinEOMsTmn}, and hence the right-hand side of eq.~\eqref{EinsteinEOMs}, is constant. This constitutes a useful simplification which, in conjunction with Eq.~\eqref{betaMetric2}, allows us to avoid solving non-linear partial differential equations and instead focus on transverse spaces -- such as $\cM_6$ in the present case -- that can be expressed as direct products of Einstein manifolds. The same strategy will be employed in constructing the RR ansätze in all examples throughout the text.

By inspecting eq. \eqref{RRformsEOM}, in view of eq. \eqref{conditionsCS2lambda}, we can easily see that the equation for $F_2$ is trivially satisfied. On the other hand, the equation for $F_4$ suggests that
\begin{subequations}\label{dconstraintsM6CS2CS2_4}\begin{align}
    & de^{01} = 0 \quad\text{if at least one between}\quad c_5,c_6,c_7,c_8 \quad\text{is not zero;} \\
    & de^{23} = 0 \quad\text{if at least one between}\quad c_9,c_{10},c_{11},c_{12} \quad\text{is not zero;} \\
    & de^{45} = 0 \quad\text{if at least one between}\quad c_{13},c_{14},c_{15},c_{16} \quad\text{is not zero.}
\end{align}\end{subequations}
Similarly, the equation for $F_6$ implies that
\begin{subequations}\label{dconstraintsM6CS2CS2_6}\begin{align}
    & de^{2345} = 0 \quad\text{if at least one between}\quad c_5,c_6,c_7,c_8 \quad\text{is not zero;} \\
    & de^{0145} = 0 \quad\text{if at least one between}\quad c_9,c_{10},c_{11},c_{12} \quad\text{is not zero;} \\
    & de^{0123} = 0 \quad\text{if at least one between}\quad c_{13},c_{14},c_{15},c_{16} \quad\text{is not zero.}
\end{align}\end{subequations}
Finally, from the equation for $F_8$ we understand that
\begin{equation}\label{dconstraintsM6CS2CS2_8}
    de^{012345}=0 \;,
\end{equation}
when any of the $c_1 , \dots , c_4$ is non-zero. Notice that the combination of eqs.~\eqref{dconstraintsM6CS2CS2_4} and \eqref{dconstraintsM6CS2CS2_6} also implies eq.~\eqref{dconstraintsM6CS2CS2_8}. Therefore, eq.~\eqref{dconstraintsM6CS2CS2_8} holds whenever any of the RR forms are non-vanishing. Below we will find examples where each of the conditions above are satisfied.

In addition, the field equations \eqref{fieldeqnH} for the NS three-form implies
\begin{subequations}\label{CconstraintsM6CS2CS2_1}\begin{align}
    &c_8c_9 - c_7c_{10} - c_6c_{11} + c_5c_{12} - c_1c_{13} - c_2c_{14} - c_3c_{15} - c_4c_{16} = 0 \;, \\
    &c_1c_9 + c_2c_{10} + c_3c_{11} + c_4c_{12} - c_8c_{13} + c_7c_{14} + c_6c_{15} - c_5c_{16} = 0 \;, \\
    &c_1c_5 + c_2c_6 + c_3c_7 + c_4c_8 + c_{12}c_{13} - c_{11}c_{14} - c_{10}c_{15} + c_9c_{16} = 0 \;.
\end{align}\end{subequations}

By combining the Einstein equations \eqref{EinsteinEOMs}, \eqref{EinsteinEOMsTmn} and the identity \eqref{betaMetric2}, we find that the components of the Ricci tensor\footnote{For convenience we work in the basis $e^a \, (a = 0 , \ldots , 9)$, and therefore the indices $a, \, b$ are flat.}
on $\cM_6$ are given in terms of the parameters $c_i$ as follows
\begin{fleqn}\begin{subequations}\label{curvaturesM6CS2CS2}\begin{equation}
    R_{ab} = - \big( c^2_1 + \ldots + c^2_{16} \big) \eta_{ab} \eqdef -r_1\eta_{ab} \;, \quad\quad\;\, a, b = 0, 1 \;,
\end{equation}\begin{equation}\begin{split}
    R_{ab} &= \big( c^2_5 + c^2_6 + c^2_7 + c^2_8 + c^2_9 + c^2_{10} + c^2_{11} + c^2_{12} - c^2_1 - c^2_2 - c^2_3 - c^2_4 - c^2_{13} - c^2_{14} \\
    &- c^2_{15} - c^2_{16} \big) \delta_{ab} \eqdef r_2\delta_{ab} \;, \qquad\qquad\qquad\quad\;\, a, b = 2, 3 \;,
\end{split}\end{equation}\begin{equation}\begin{split}
    R_{ab} &= \big( c^2_5 + c^2_6 + c^2_7 + c^2_8 + c^2_{13} + c^2_{14} + c^2_{15} + c^2_{16} - c^2_1 - c^2_2 - c^2_3 - c^2_4 - c^2_9 - c^2_{10} \\ 
    &- c^2_{11} - c^2_{12} \big) \delta_{ab} \eqdef r_3\delta_{ab} \;, \qquad\qquad\qquad\quad\;\, a, b = 4, 5 \;.
\end{split}\end{equation}\end{subequations}\end{fleqn}
The rest of the Einstein equations yield the following constraints
\begin{fleqn}\begin{subequations}\label{CconstraintsM6CS2CS2_2}\begin{equation}\begin{split}
    \mu &= c^2_1 + c^2_2 - c^2_3 - c^2_4 - c^2_5 - c^2_6 + c^2_7 + c^2_8 + c^2_9 + c^2_{10} - c^2_{11} - c^2_{12} + c^2_{13} + c^2_{14} \\
    & - c^2_{15} - c^2_{16} \;,
\end{split}\end{equation}\begin{equation}\begin{split}
    \mu &= c^2_1 - c^2_2 + c^2_3 - c^2_4 - c^2_5 + c^2_6 - c^2_7 + c^2_8 + c^2_9 - c^2_{10} + c^2_{11} - c^2_{12} + c^2_{13} - c^2_{14} \\
    &+ c^2_{15} - c^2_{16} \;,
\end{split}\end{equation}\begin{equation}\begin{split}
    0 &= c_1c_3 + c_2c_4 - c_5c_7 - c_6c_8 + c_9c_{11} + c_{10}c_{12} + c_{13}c_{15} + c_{14}c_{16} \;,
\end{split}\end{equation}\begin{equation}\begin{split}
    0 &= c_1c_2 + c_3c_4 - c_5c_6 - c_7c_8 + c_9c_{10} + c_{11}c_{12} + c_{13}c_{14} + c_{15}c_{16} \;.
\end{split}\end{equation}\end{subequations}\end{fleqn}
The structure of the Ricci tensor \eqref{curvaturesM6CS2CS2}, together with the conditions \eqref{dconstraintsM6CS2CS2_4}, \eqref{dconstraintsM6CS2CS2_6}, and \eqref{dconstraintsM6CS2CS2_8}, indicates whether the six-dimensional space $\cM_6$ splits into a direct product of lower-dimensional submanifolds. In particular, we will construct examples corresponding to some of the following cases:
\begin{longtable}{|c|c|c|}
    \hline
    \textbf{Split of $\cM_6$} & \textbf{Spacetime directions} & \textbf{Conditions} \\
    \hline\hline
    $\cM^t_6$ & $\cM^t_6: \quad e^0 , \ldots , e^5$ &   $de^{012345}=0 \quad \& \quad r_1 = -r_2 = -r_3$ \\
    \hline
    \multirow{2}{*}{$\cM^t_4 \times \cM_2$} & $\cM^t_4: \quad e^0 , \ldots , e^3$ & \multirow{2}{*}{$de^{0123} = de^{45} = 0 \quad \& \quad r_1 = - r_2$} \\
     & $\cM_2: \quad e^4 , \, e^5$ & \\
    \hline
    \multirow{2}{*}{$\cM^t_2 \times \cM_4$} & $\cM^t_2: \quad e^0 , \, e^1$ & \multirow{2}{*}{$de^{01} = de^{2345} = 0 \quad \& \quad r_2 = r_3$} \\
     & $\cM_4: \quad e^2 , \ldots , e^5$ & \\
    \hline
    \multirow{3}{*}{$\cM^t_2 \times \cM_2 \times \cM'_2$} & $\cM^t_2: \quad e^0 , \, e^1$ & \multirow{3}{*}{$de^{01} = de^{23} = de^{45} = 0$} \\
     & $\cM_2: \quad e^2 , \, e^3$ & \\
     & $\cM'_2: \quad e^4 , \, e^5$ &\\
     \hline
\caption{Decomposition of $\cM_6$ into subspaces and the associated conditions.}
    \label{Tab:SplitM6}
\end{longtable}
\noindent In the above, we denote the space containing the time direction with the superscript $t$. The conditions in the third column translate into setting certain $c_i$ coefficients to zero. For instance, imposing $de^{012345} = 0$, without requiring either of eqs.~\eqref{dconstraintsM6CS2CS2_4} or \eqref{dconstraintsM6CS2CS2_6}, implies that $c_5 = \ldots = c_{16} = 0$, which in turn ensures that $r_1 = - r_2 = - r_3$. Conversely, requiring $r_1 = - r_2 = - r_3$ directly also leads to $c_5 = \ldots = c_{16} = 0$. As we will see, in this case $\cM^t_6$ can be taken to be $\rm AdS_6$ or a direct product of a lower-dimensional $\rm AdS$ space with hyperbolic spaces. Analogous reasoning applies to the remaining examples.

Notice also that the Ricci tensor \eqref{curvaturesM6CS2CS2} suggests that the various subspaces are Einstein manifolds, where $r_1 > 0$ and
\begin{equation}\label{rconstraintsM6CS2CS2}
    -r_1 + r_2 + r_3 = -\nu \, . 
\end{equation}
The latter is implied by eqs. \eqref{curvatureM6} and \eqref{curvaturesM6CS2CS2}. In the following we are going to solve the constraints \eqref{CconstraintsM6CS2CS2_1}, \eqref{curvaturesM6CS2CS2}, \eqref{CconstraintsM6CS2CS2_2} and \eqref{rconstraintsM6CS2CS2} in some specific cases\footnote{The equations \eqref{CconstraintsM6CS2CS2_1}, \eqref{curvaturesM6CS2CS2}, \eqref{CconstraintsM6CS2CS2_2} and \eqref{rconstraintsM6CS2CS2} provide $11$ independent conditions for a total of $19$ parameters. This indicates a large degree of freedom, which we do not intend to fully exploit. Although the symmetries of these equations could in principle be used to restrict the parameter space, in our approach we instead fix a subset of parameters to zero according to the specific decomposition of the external space that we seek to realise in each case.}.

\subsubsection*{Example 1: $\cM_6 \simeq {\rm AdS_6, \; AdS_4 \times H_2, \;AdS_3\times H_3,\; AdS_2 \times H_4, \; AdS_2 \times H_2 \times H'_2}$}
\label{Example3111}

Let us start with the assumption of a vanishing $F_4$, that is to say by setting $c_5 = c_6 = \dots = c_{16} = 0$. Such a choice is justified below Table~\ref{Tab:SplitM6}. In this case, eqs. \eqref{CconstraintsM6CS2CS2_1} are trivially satisfied, while from eqs. \eqref{curvaturesM6CS2CS2} we get that
\begin{equation}
    r_1 = - r_2 = - r_3 = c^2_1 + c^2_2 + c^2_3 + c^2_4 \;.
\end{equation}
In other words, the Ricci tensor on $\cM_6$ takes the form
\begin{equation}
    R_{ab} = - r_1\eta_{ab}\;, \quad a,b = 0,\dots,5\;,
\end{equation}
which allows us to consider any of the following options: ${\rm AdS_6}$, ${\rm AdS_4 \times H_2}$, ${\rm AdS_3\times H_3}$, ${\rm AdS_2 \times H_4}$ or ${\rm AdS_2 \times H_2 \times H'_2}$. The rest of the constraints now read as follows
\begin{subequations}\begin{align}
    & \mu = c^2_1 + c^2_2 - c^2_3 - c^2_4 = c^2_1 - c^2_2 + c^2_3 - c^2_4 \;, \\
    & 0 = c_1c_3 + c_2c_4 \;, \\
    & 0 = c_1c_2 + c_3c_4 \;, \\
    & r_1 = \frac\nu 3 \;.
\end{align}\end{subequations}
One way to solve the above equations is to take
\begin{tcolorbox}[colback=blue!50!green!5,colframe=black,title={},width=1.01\textwidth,left=0pt, right=0pt]\begin{align}
    c_1 = s_1 \sqrt{\frac\nu6 + \frac\mu2} \;, \qquad
    c_4 = s_4 \sqrt{\frac\nu6 - \frac\mu2} \;,
\end{align}\end{tcolorbox}
\noindent with $s_i = \pm 1$ \footnote{The different sign choices reflect the symmetry of the supergravity equations of motion under $F_p \to - F_p$, which in this example applies separately to each component of the RR fields. Such choices are not always innocent; for instance, in supersymmetric backgrounds, they may affect the amount of supersymmetry that is preserved.}, and keep the rest of the $c_i$ zero. Imposing reality on the RR forms restricts the allowed values of $\lambda$. Indeed, this leads to considering $\nu \geq 3 \mu$, which translates to
\begin{equation}
    0 \leq \lambda \leq \frac{1}{2} \big( 3 - \sqrt{5} \big) \;.
\end{equation}
The above bound excludes the non-Abelian T-dual limit $\lambda \to 1$.

\subsubsection*{Example 2: $\cM_6 \simeq {\rm AdS_4 \times (S^2, H_2)}$}
\label{Example3112}

Aiming for a decomposition of the form $\cM_6 \simeq {\rm AdS_4} \times \cM_2$, we are instructed by eqs. \eqref{dconstraintsM6CS2CS2_4}, \eqref{dconstraintsM6CS2CS2_6} to set $c_5 = \dots = c_{12} = 0$. The rationale behind this choice is to avoid imposing the constraints 
$de^{01} = de^{23} = de^{2345} = de^{0145} = 0$, 
which are incompatible with the aforementioned decomposition of $\cM_6$. 
Alternatively, in order to realise $\cM_6 \simeq {\rm AdS}_4 \times \cM_2$, 
one must require $r_1 = - r_2$ in eq.~\eqref{curvaturesM6CS2CS2}, 
which again leads to setting $c_5 , \, \ldots \, , c_{12}$ to zero. The Ricci tensor on $\cM_6$ now splits into the following two parts
\begin{subequations}\begin{align}
    R_{ab} &= -r_1\eta_{ab} \;, \qquad\;\, a, b = 0, 1, 2, 3 \;,
    \label{curvaturesM6CS2CS2Ex2AdS4}\\
    R_{ab} &= r_3\delta_{ab} \;, \qquad\quad\,\, a, b = 4, 5 \;, \label{curvaturesM6CS2CS2Ex2S2}
\end{align}\end{subequations}
where
\begin{subequations}\begin{align}
    r_1 &= - r_2 = c^2_1 + c^2_2 + c^2_3 + c^2_4 + c^2_{13} + c^2_{14} + c^2_{15} + c^2_{16} \;,
    \\
    r_3 &= - c^2_1 - c^2_2 - c^2_3 - c^2_4 + c^2_{13} + c^2_{14} + c^2_{15} + c^2_{16} \;.
\end{align}\end{subequations}
The remaining conditions on the parameters $c_i$ and the curvatures $r_i$ become
\begin{subequations}\begin{align}
    \mu & = c^2_1 + c^2_2 - c^2_3 - c^2_4 + c^2_{13} + c^2_{14} - c^2_{15} - c^2_{16} \;,
    \\
    \mu & = c^2_1 - c^2_2 + c^2_3 - c^2_4 + c^2_{13} - c^2_{14} + c^2_{15} - c^2_{16} \; ,
    \\
    0 &= c_1c_3 + c_2c_4 + c_{13}c_{15} + c_{14}c_{16} \; ,
    \\
    0 &= c_1c_2 + c_3c_4 + c_{13}c_{14} + c_{15}c_{16} \; ,
    \\
    \nu & = 2 r_1 - r_3 \; .
\end{align}\end{subequations}

One possibility is to look for solutions with $r_3 > 0$. An obvious choice would be to switch-off the RR two-form leading to
\begin{tcolorbox}[colback=blue!50!green!5,colframe=black,title={},width=1.01\textwidth,left=0pt, right=0pt]\begin{equation}\begin{gathered}
    c_{13} = s_{13} \sqrt{\frac{\nu + \mu}{2}} \; , \qquad
    c_{16} = s_{16} \sqrt{\frac{\nu - \mu}{2}} \; , \qquad
    r_1 = r_3 = \nu \;,
\end{gathered}\end{equation}\end{tcolorbox}
\noindent with $s_i = \pm 1$ and the rest of the $c_i$ parameters being zero. This solution is real for all values of $\lambda$ in the fundamental domain $[0,1)$. Moreover, $r_1, \; r_3 > 0$ for $\lambda \in [0 , 1)$. Therefore, we can safely choose $\cM_6 \simeq \rm{AdS_4 \times S^2}$, where the ${\rm AdS_4}$ is normalised according to \eqref{curvaturesM6CS2CS2Ex2AdS4} and $S^2$ as in \eqref{curvaturesM6CS2CS2Ex2S2}.

A second solution can be found by keeping only $c_2$, $c_3$ and $c_{13}$ to be non-zero. In particular, we set
\begin{tcolorbox}[colback=blue!50!green!5,colframe=black,title={},width=1.01\textwidth,left=0pt, right=0pt]\begin{equation}\begin{gathered}
    c_2 = s_2 \sqrt{\frac\nu6 - \frac\mu6} \;, \quad
    c_3 = s_3 \sqrt{\frac\nu6 - \frac\mu6} \;, \quad
    c_{13} = s_{13} \sqrt{\mu} \;, \\
    r_1 = \frac13 (2\mu+\nu) \;, \quad
    r_3 = \frac13 (4\mu-\nu)\;,
\end{gathered}\end{equation}\end{tcolorbox}
\noindent where $s_i$ are signs. The reality of $c_2$ and $c_3$, together with positiveness of $r_3$, imply that $\lambda$ should be restricted in the range $[2-\sqrt{3}, 1)$. In this domain we can interpret $\cM_2$ as $\rm S^2$. Notice that the conformal point ($\lambda=0$) is excluded from this interval. Moreover, in the case of $\lambda = 2-\sqrt{3}$, we have $r_3 = 0$. Therefore, for this specific value, $\cM_2$ corresponds to a flat space. In contrary, the request for a negative curvature $r_3\leq0$ subspace $\cM_2$ would impose a different bound for $\lambda$, in particular $[0, 2-\sqrt{3}]$. In this scenario, the non-Abelian T-dual limit ($\lambda\to1$) is excluded and the space $\cM_2$ can be safely chosen to be $\rm H_2$.

\subsubsection*{Example 3: $\cM_6 \simeq {\rm AdS_2 \times (T^4, \mathbb{CP}^2, S^4, S^2 \times S'^2)}$}
\label{Example3113}

Assuming $c_9 = c_{10} = \dots = c_{16} = 0$, we get manifold of the kind ${\rm AdS_2 \times} \cM_4$, as it can be understood from eqs. \eqref{dconstraintsM6CS2CS2_4}, \eqref{dconstraintsM6CS2CS2_6} and \eqref{dconstraintsM6CS2CS2_8}. Indeed, this choice allows us to avoid imposing the conditions $de^{23} = de^{45} = de^{0145} = de^{0123} = 0$. The Ricci tensor $R_{ab}$ on $\cM_6$ can be decomposed as
\begin{subequations}\label{RicciEx33}\begin{align}
    &R_{ab} = -r_1\eta_{ab} \;, \qquad\;\, a,b = 0,1 \;, \\
    &R_{ab} = r_2\delta_{ab} \;, \qquad\quad\; a,b = 2,3,4,5 \;,
\end{align}\end{subequations}
with
\begin{subequations}\begin{align}
    r_1 &= c^2_1 + c^2_2 + c^2_3 + c^2_4 + c^2_5 + c^2_6 + c^2_7 + c^2_8 \;,
    \\
    r_2 &= r_3 = - c^2_1 - c^2_2 - c^2_3 - c^2_4 + c^2_5 + c^2_6 + c^2_7 + c^2_8 \;.
\end{align}\end{subequations}
The other conditions on the curvatures $r_i$ and the parameters $c_i$ now read
\begin{subequations}\begin{align}
    \mu & = c^2_1 + c^2_2 - c^2_3 - c^2_4 - c^2_5 - c^2_6 + c^2_7 + c^2_8 \;,
    \\
    \mu & = c^2_1 - c^2_2 + c^2_3 - c^2_4 - c^2_5 + c^2_6 - c^2_7 + c^2_8 \; ,
    \\
    0 &= c_1c_3 + c_2c_4 - c_5c_7 - c_6c_8 \; ,
    \\
    0 &= c_1c_2 + c_3c_4 - c_5c_6 - c_7c_8 \; ,
    \\
    0 &= c_1c_5 + c_2c_6 + c_3c_7 + c_4c_8 \; ,
    \\
    \nu & = r_1 - 2r_2 \; .
\end{align}\end{subequations}
Among the many possibilities that we have, we can choose $c_1 = c_4 = c_6 = c_7 = 0$, giving the solution
\begin{tcolorbox}[colback=blue!50!green!5,colframe=black,title={},width=1.01\textwidth,left=0pt, right=0pt]\begin{equation}\begin{gathered}
    c_2 = s_2 \sqrt{\frac{\mu + \nu}{6} + \frac{c^2_5}{3}} \;, \quad
    c_3 = s_3 \sqrt{\frac{\mu + \nu}{6} + \frac{c^2_5}{3}} \;, \quad
    c_8 = s_8 \sqrt{\mu + c^2_5} \;, \\
    r_1 = \frac{1}{3} \big( 4 \mu + \nu + 8 c^2_5 \big) \;, \quad
    r_2 = \frac{1}{3} \big( 2 \mu - \nu + 4 c^2_5 \big) \;,
\end{gathered}\end{equation}\end{tcolorbox}
\noindent with $s_i = \pm 1$ and $c_5$ being a free parameter. The solution is real for all values of $\lambda$ in $[0 , 1)$, and $r_2 \geq 0$ provided that
\begin{equation}
    c^2_5 \geq \frac{\nu - 2 \mu}{4} \geq 0 \;.
\end{equation}
Therefore, provided that $c_5^2 > \frac{\nu - 2\mu}{4}$,
the space $\cM_4$ can be chosen to be a four-sphere $\rm S^4$, the product space $\rm S^2 \times S'^2$, or $\mathbb{CP}^2$. In the special case where $c_5^2 = \frac{\nu - 2\mu}{4}$, $\cM_4$ may instead be taken to be a four-torus $\rm T^4$.

\subsubsection*{Example 4: $\cM_6 \simeq {\rm AdS_2 \times \mathbb{CP}^2}$}
\label{Example3114}

As a last example for this class of backgrounds, we would like to examine the case where $c_{13} = -c_9$, $c_{14} = - c_{10}$, $c_{15} = - c_{11}$ and $c_{16} = - c_{12}$. With this in mind, we see that the equation \eqref{RRformsEOM} for $F_4$ now suggests
\begin{subequations}\begin{align}
    & de^{01} = 0 \qquad\qquad\,\, \text{if at least one between}\quad c_5,c_6,c_7,c_8 \quad\text{is not zero;} \\
    & d \big( e^{23} - e^{45} \big) = 0 \quad\text{if at least one between}\quad c_9,c_{10},c_{11},c_{12} \quad\text{is not zero,} \label{dconstraintsM6CS2CS2_4_CP2}
\end{align}\end{subequations}
instead of \eqref{dconstraintsM6CS2CS2_4}. Similarly, the equation for $F_6$ implies
\begin{subequations}\begin{align}
    & de^{2345} = 0 \qquad\qquad\qquad\quad\, \text{if at least one between}\quad c_5,c_6,c_7,c_8 \quad\text{is not zero;} \\
    & d\left( e^{01} \wedge \big( e^{23} - e^{45} \big) \right) = 0 \quad\text{if at least one between}\quad c_9,c_{10},c_{11},c_{12} \quad\text{is not zero,}
\end{align}\end{subequations}
instead of \eqref{dconstraintsM6CS2CS2_6}. On the other hand, \eqref{dconstraintsM6CS2CS2_8} still remains true.

As for the Ricci tensor, it still takes the form \eqref{RicciEx33}, but now the values of the curvatures $r_1$ and $r_2$ are given by
\begin{subequations}\begin{align}
    r_1 &= c^2_1 + c^2_2 + c^2_3 + c^2_4 + c^2_5 + c^2_6 + c^2_7 + c^2_8 + 2c^2_9 + 2c^2_{10} + 2c^2_{11} + 2c^2_{12} \;,
    \\
    r_2 &= r_3 = - c^2_1 - c^2_2 - c^2_3 - c^2_4 + c^2_5 + c^2_6 + c^2_7 + c^2_8 \;.
\end{align}\end{subequations}
From the structure of the six-dimensional space one can suspect that it is going to break into the direct product of two Einstein submanifolds, of dimension two and four respectively, where the two-dimensional one has negative curvature.
For the other constant parameters that enter the RR parametrisations, we find that they should satisfy the following relations
\begin{subequations}\begin{align}
    \mu &= c_1^2 + c_2^2 - c_3^2 - c_4^2 - c_5^2 - c_6^2 + c_7^2 + c_8^2 + 2 c_9^2 + 2 c_{10}^2 - 2 c_{11}^2 - 2 c_{12}^2 \;, \\
    \mu &= c_1^2 - c_2^2 + c_3^2 - c_4^2 - c_5^2 + c_6^2 - c_7^2 + c_8^2 + 2 c_9^2 - 2 c_{10}^2 + 2 c_{11}^2 - 2 c_{12}^2 \;, \\
    0 &= c_1 c_9 + c_8 c_9 + c_2 c_{10} - c_7 c_{10} + c_3 c_{11} - c_6 c_{11} + c_4 c_{12} + c_5 c_{12} \;, \\
    0 &= c_1 c_5 + c_2 c_6 + c_3 c_7 + c_4 c_8 + 2 c_{10} c_{11} - 2 c_9 c_{12} \;, \\
    0 &= c_1 c_3 + c_2 c_4 - c_5 c_7 - c_6 c_8 + 2 c_9 c_{11} + 2 c_{10} c_{12} \;, \\
    0 &= c_1 c_2 + c_3 c_4 - c_5 c_6 - c_7 c_8 + 2 c_9 c_{10} + 2 c_{11} c_{12} \;, \\
    \nu &= r_1 - 2 r_2 \;.
\end{align}\end{subequations}
The subsequent choice of parameters
\begin{tcolorbox}[colback=blue!50!green!5,colframe=black,title={},width=1.01\textwidth,left=0pt, right=0pt]\begin{equation}\begin{gathered}
    c_4 = s_4 \frac{\mu + \nu}{\sqrt{8\mu + 4\nu}} \;, \quad
    c_8 = s_8 \frac{3\mu + \nu}{\sqrt{8\mu + 4\nu}} \;, \quad
    c_{10} = s_{10} \frac{\sqrt{3\mu^2 + 4\mu\nu + \nu^2}}{2\sqrt{8\mu + 4\nu}} \;, \\
    c_{11} = s_{11} \frac{\sqrt{3\mu^2 + 4\mu\nu + \nu^2}}{2\sqrt{8\mu + 4\nu}} \;, \quad
    r_1 = 2\mu + \nu \;, \quad
    r_2 = \mu \;,
\end{gathered}\end{equation}\end{tcolorbox}
\noindent with the remaining being trivial, provides a real solution to the equations. Furthermore, $s_i=\pm1$ and satisfy $s_4s_8 + s_{10}s_{11} = 0$. The deformation parameter $\lambda$ takes values in $[0,1)$. The Ricci tensor on $\cM_6$, given in eq.~\eqref{RicciEx33}, suggests that the subspaces spanned by $(e^0 , e^1)$ and $(e^2 , e^3 , e^4 , e^5)$ are Einstein with negative and positive curvature, respectively. In addition, the condition \eqref{dconstraintsM6CS2CS2_4_CP2} requires the two-form $e^{23} - e^{45}$ to be closed. These properties justify the choice $\cM_6 = {\rm AdS}_2 \times \mathbb{CP}^2$, with $\mathbb{CP}^2$ extending along $(e^2 , e^3 , e^4 , e^5)$ and $e^{23} - e^{45}$ identified as its K\"ahler form (see Appendix~\ref{Sec:Appendix_Single_CS2s} for further details).


\refstepcounter{subsubsection}\subsubsection*{\thesubsubsection\quad Type-IIA on $\cM_4 \times {\rm CS}^2_\lambda \times {\rm CS}^2_\lambda \times {\rm CS}^2_\lambda$}
\label{Subsubsec:IIA_M4CS2CS2CS2}

A straightforward generalisation of the previous setup is to consider three copies of the $\rm CS^2_\lambda$ model (see Sec. \ref{Subsec:CS^2_Model}). From the ten-dimensional point of view, the geometry will be given by the direct product
\begin{equation}
    \cM_4 \times \rm CS^2_\lambda \times CS^2_\lambda \times CS^2_\lambda \;.
\end{equation}
Again, the time direction has to be included in the subspace $\cM_4$, which is yet to be determined, and whose line-element can be expressed by means of the frame $(e^0,\dots,e^3)$ as
\begin{equation}
    ds^2_{\cM_4} = - (e^0)^2 + (e^1)^2 + (e^2)^2 + (e^3)^2 \;.
\end{equation}
The remaining six dimensions are characterised by a Euclidean signature and are spanned by the frame fields $e^4 \to \mathfrak{e}^1$, $e^5 \to \mathfrak{e}^2$; $e^6 \to \mathfrak{e}^1$, $e^7 \to \mathfrak{e}^2$; and $e^8 \to \mathfrak{e}^1$, $e^9 \to \mathfrak{e}^2$, with $\mathfrak{e}^1 , \, \mathfrak{e}^2$ given in \eqref{Eq:frame2}. In order to differentiate between the three copies of $\rm CS^2_\lambda$, we are going to adopt three different sets of coordinates $(x_1,\omega_1)$, $(x_2,\omega_2)$ and $(x_3,\omega_3)$. We keep the assumption of a vanishing NS two-form, and take the dilaton to be the sum of scalars characterising each copy of $\rm CS^2_\lambda$
\begin{equation}\label{Dilaton_222}
    \Phi = - \frac12 \log \left(\frac{2\omega_1^2}{\omega_{1+}^2}\right) - \frac12 \log \left(\frac{2\omega_2^2}{\omega_{2+}^2}\right) - \frac12 \log \left(\frac{2\omega_3^2}{\omega_{3+}^2}\right) \;.
\end{equation}

From the dilaton eq. \eqref{dilatonEOM} and \eqref{betaScalar2}, one can infer the curvature of the four-dimensional manifold $\cM_4$
\begin{equation}\label{curvatureM4_222}
    R_{\cM_4} = -3\nu \;.
\end{equation}
This is again constant and negative for all values of the deformation parameter $\lambda\in[0,1)$.

The geometric properties of $\cM_4$ can be further inspected by providing an explicit ansatz for the RR forms. We propose such an ansatz following the reasoning discussed in Sec.~\ref{Subsubsec:IIA_M6CS2CS2}, below Eq.~\eqref{RRansatzEx311}. In particular, we construct a form that:
\begin{enumerate}
    \item Lives entirely in the internal space, which in this case is ${\rm CS}^2_{\lambda} \times {\rm CS}^2_{\lambda} \times {\rm CS}^2_{\lambda}$,  
    \item Is closed, and  
    \item Has a Hodge dual (taken within ${\rm CS}^2_{\lambda} \times {\rm CS}^2_{\lambda} \times {\rm CS}^2_{\lambda}$) that is also closed.  
\end{enumerate}
For this purpose, we exploit the property in eq.~\eqref{conditionsCS2lambda}. In this example, such a form can be expressed as a linear combination of $e^{-\Phi} e^{468}$, $e^{-\Phi} e^{469}$, $e^{-\Phi} e^{478}$, $e^{-\Phi} e^{479}$, $e^{-\Phi} e^{568}$, $e^{-\Phi} e^{569}$, $e^{-\Phi} e^{578}$ and $e^{-\Phi} e^{579}$ with the important feature that the relative coefficients are constants. If any of the three legs were missing in any of these components, the resulting two-form would not be closed, because the dilaton is the sum of all three scalars of each ${\rm CS}^2_{\lambda}$ model. Therefore, the form we are seeking in this case is of rank three, and it is this form that we use to construct the RR fluxes, which must have at least rank three. As the rank of the two-form $F_2$ is not large enough to accommodate legs that sit in all three copies of ${\rm CS}^2_\lambda$ at the same time, we will assume that it is trivial. Therefore, the only available option is to turn on the four-form $F_4$. More precisely, we propose the ansatz
\begin{equation}\begin{split}
    F_4 & = 2 e^{-\Phi} e^{2} \wedge \big( c_1 e^{468} + c_2 e^{469} + c_3 e^{478} + c_4 e^{479} \\
    & \hspace{1.8cm} + c_5 e^{568} + c_6 e^{569} + c_7 e^{578} + c_8 e^{579} \big) \\
    & +\, 2 e^{-\Phi} e^{3} \wedge \big( c_9 e^{468} + c_{10} e^{469} + c_{11} e^{478} + c_{12} e^{479} \\
    &\hspace{1.8cm} + c_{13} e^{568} + c_{14} e^{569} + c_{15} e^{578} + c_{16} e^{579} \big) \;,
\end{split}\end{equation}
with $c_i \, (i = 1 , \dots , 16)$ being constants.

The equation of motion \eqref{RRformsEOM} for the RR four-form, together with the relation \eqref{conditionsCS2lambda}, implies that
\begin{subequations}\label{dconstraintsM4CS2CS2CS2_4}\begin{align}
    & de^{2} = 0 \quad\text{if at least one between}\quad c_1, \dots, c_8 \quad\text{is not zero;} \\
    & de^{3} = 0 \quad\text{if at least one between}\quad c_9, \dots, c_{16} \quad\text{is not zero.}
\end{align}\end{subequations}
In the same way, the field equation for the RR six-form suggests that
\begin{subequations}\label{dconstraintsM4CS2CS2CS2_6}\begin{align}
    & de^{012} = 0 \quad\text{if at least one between}\quad c_9, \dots, c_{16} \quad\text{is not zero;} \\
    & de^{013} = 0 \quad\text{if at least one between}\quad c_1, \dots, c_8 \quad\text{is not zero.}
\end{align}\end{subequations}
Moreover, the equation of motion \eqref{fieldeqnH} for the NS three-form $H$ gives
\begin{equation}\label{CconstraintsM4CS2CS2CS2_1}
    c_1c_{16} - c_2c_{15} - c_3c_{14} + c_4c_{13} - c_5c_{12} + c_6c_{11} + c_7c_{10} - c_8c_9 = 0 \;.
\end{equation}

The Einstein equations \eqref{EinsteinEOMs}, \eqref{EinsteinEOMsTmn} imply that the non-vanishing components of the Ricci tensor on $\cM_4$ take the form
\begin{subequations}\label{curvaturesM4CS2CS2CS2_IIA}\begin{align}
    R_{ab} &= -\left( c_1^2 + \dots + c_{16}^2 \right) \eta_{ab} \eqdef -r_1\eta_{ab} \;, \hspace{3.55cm} a,b = 0,1 \;, \\
    R_{22} &= -R_{33} = c_1^2 + \dots + c_{8}^2 - c_9^2 - \dots - c_{16}^2 \;, \\
    R_{23} &= R_{32} = 2\,(c_1 c_9+c_2 c_{10}+c_3 c_{11}+c_4 c_{12}+c_5 c_{13}+c_6 c_{14}+c_7 c_{15}+c_8 c_{16}) \;.
\end{align}\end{subequations}
The remaining Einstein equations, when combined with eq. \eqref{betaMetric2}, give the following constraints
\begin{fleqn}\begin{subequations}\label{CconstraintsM4CS2CS2CS2_2}\begin{equation}
    0 = c_8c_9 - c_7c_{10} - c_6c_{11} + c_5c_{12} - c_4c_{13} + c_3c_{14} + c_2c_{15} - c_1c_{16} \;,
\end{equation}\begin{equation}
    0 = c_1 c_5+c_2 c_6+c_3 c_7+c_4 c_8+c_9 c_{13}+c_{10} c_{14}+c_{11} c_{15}+c_{12} c_{16} \;,
\end{equation}\begin{equation}
    0 = c_1 c_3+c_2 c_4+c_5 c_7+c_6 c_8+c_9 c_{11}+c_{10} c_{12}+c_{13} c_{15}+c_{14} c_{16} \;,
\end{equation}\begin{equation}\begin{split}
    0 &= c_1 c_2+c_3 c_4+c_5 c_6+c_7 c_8+c_9 c_{10}+c_{11} c_{12}+c_{13} c_{14}+c_{15} c_{16} \;,
\end{split}\end{equation}\begin{equation}\begin{split}
    \mu &= c_1^2+c_2^2+c_3^2+c_4^2-c_5^2-c_6^2-c_7^2-c_8^2+c_9^2+c_{10}^2+c_{11}^2+c_{12}^2 \\
    &-c_{13}^2-c_{14}^2-c_{15}^2-c_{16}^2 \;,
\end{split}\end{equation}\begin{equation}\begin{split}
    \mu &= c_1^2+c_2^2-c_3^2-c_4^2+c_5^2+c_6^2-c_7^2-c_8^2+c_9^2+c_{10}^2-c_{11}^2-c_{12}^2 \\
    &+c_{13}^2+c_{14}^2-c_{15}^2-c_{16}^2 \;,
\end{split}\end{equation}\begin{equation}\begin{split}
    \mu &= c_1^2-c_2^2+c_3^2-c_4^2+c_5^2-c_6^2+c_7^2-c_8^2+c_9^2-c_{10}^2+c_{11}^2-c_{12}^2 \\
    &+c_{13}^2-c_{14}^2+c_{15}^2-c_{16}^2 \;.
\end{split}\end{equation}\end{subequations}\end{fleqn}

By inspecting eqs. \eqref{dconstraintsM4CS2CS2CS2_4} it is natural to assume that $\cM_4$ factorises as $\cM_4 = \cM^t_2 \times {\rm T^2}$, with $\cM^t_2$ spanned by $(e^0 , \, e^1)$ and the torus being extended in the $(e^2 , \,  e^3)$ directions. In this case, the Ricci tensor components $R_{22}, \, R_{33}, \, R_{23}$ and $R_{32}$ vanish, providing two additional constraints. If we further combine \eqref{curvaturesM4CS2CS2CS2_IIA} with \eqref{curvatureM4_222} we get
\begin{equation}\label{r1M4CS2CS2CS2_IIA}
    r_1 = \frac32 \nu \;.
\end{equation}
A solution to the above system of equations is given by
\begin{tcolorbox}[colback=blue!50!green!5,colframe=black,title={},width=1.01\textwidth,left=0pt, right=0pt]\begin{equation}\label{SolSec312}\begin{gathered}
    c_2 = s_2 \sqrt{\frac{\nu}{4} + \frac{\mu}{6}} \;, \quad
    c_7 = s_7 \sqrt{\frac{\nu}{2} - \frac{\mu}{6}} \;, \quad
    c_9 = s_9 \sqrt{\frac{\nu}{4} + \frac{2 \mu}{3}} \;, \\[5pt]
    c_{12} = s_{12} \sqrt{\frac{\nu}{4} - \frac{\mu}{3}} \;, \quad
    c_{14} = s_{14} \sqrt{\frac{\nu}{4} - \frac{\mu}{3}} \;.
\end{gathered}\end{equation}\end{tcolorbox}
\noindent with the rest of the $c_i$ parameters being zero and the $s_i$ representing signs. Such solution is real for all values of $\lambda$ in $[0,1)$. Moreover, given the form of the Ricci tensor \eqref{curvaturesM4CS2CS2CS2_IIA} together with \eqref{r1M4CS2CS2CS2_IIA}, we observe that $\cM^t_2$ is an Einstein space with constant negative curvature. Therefore, it is natural to interpret it as an ${\rm AdS_2}$ space, i.e. $\cM_4 = {\rm AdS_2 \times T^2}$.


\refstepcounter{subsubsection}\subsubsection*{\thesubsubsection\quad Type-IIB on $\cM_4 \times {\rm CS}^2_\lambda \times {\rm CS}^2_\lambda \times {\rm CS}^2_\lambda$}
\label{Subsubsec:IIB_M4CS2CS2CS2}

The NS sector of the preceding example \ref{Subsubsec:IIA_M4CS2CS2CS2} also supports solutions to type-IIB supergravity. The curvature on $\cM_4$ is, as usual, computed from the dilaton eq. \eqref{dilatonEOM} and is again given by eq. \eqref{curvatureM4_222}. The construction is completed by suggesting an ansatz for the type-IIB RR forms $F_1$, $F_3$, and $F_5$.

Since the form degree of $F_1$ is too small, we will take it to be zero. For the rest of the RR forms we adopt the following ansatz
\begin{subequations}\label{M4CS2CS2CS2IIBRRAnsatz}\begin{align}
    F_3 & = 2 e^{-\Phi} \big( c_1 e^{468} + c_2 e^{469} + c_3 e^{478} + c_4 e^{479} \nonumber \\
    & \hspace{1cm} + c_5 e^{568} + c_6 e^{569} + c_7 e^{578} + c_8 e^{579} \big) \;, \\
    F_5 & = 2 e^{-\Phi} (1 + \star) \, e^{23} \wedge \big( c_9 e^{468} + c_{10} e^{469} + c_{11} e^{478} + c_{12} e^{479} \nonumber \\
    & \hspace{1cm} + c_{13} e^{568} + c_{14} e^{569} + c_{15} e^{578} + c_{16} e^{579} \big) \;,
\end{align}\end{subequations}
with $c_1, \dots, c_{16}$ being constant parameters to be determined.

The field equation \eqref{RRformsEOM} for the self-dual RR five-form and the relation \eqref{conditionsCS2lambda} suggest that
\begin{equation}\label{dconstraintsM4CS2CS2CS2_5}
    de^{01} = 0 \;, \quad de^{23} = 0 \;,
\end{equation}
if at least one between $c_9, \dots, c_{16}$ is not zero.
In the same way, the equation of motion for the RR seven-form implies that
\begin{equation}\label{dconstraintsM4CS2CS2CS2_7}
    de^{0123} = 0
\end{equation}
if at least one between $c_1, \dots, c_8$ is not zero.

Furthermore, the field equations \eqref{fieldeqnH} for the NS three-form $H$ give
\begin{subequations}\label{CconstraintsM4CS2CS2CS2_1_IIB}\begin{align}
    0 &= c_1c_9 + c_2c_{10} + c_3c_{11} + c_4c_{12} + c_5c_{13}\ + c_6c_{14} + c_7c_{15} + c_8c_{16} \;, \\
    0 &= c_1c_{16} - c_2c_{15} - c_3c_{14} + c_4c_{13} - c_5c_{12} + c_6c_{11} + c_7c_{10} - c_8c_9 \;.
\end{align}\end{subequations}

The components of the Ricci tensor on $\cM_4$ are determined by the Einstein equations \eqref{EinsteinEOMs}, \eqref{EinsteinEOMsTmn}. These are expressed in terms of the parameters $c_i$ as
\begin{subequations}\label{curvaturesM4CS2CS2CS2_IIB}\begin{align}
    R_{ab} &= -\left( c_1^2 + \dots + c_{16}^2 \right) \eta_{ab} \eqdef -r_1\eta_{ab} \;, \hspace{3.55cm} a,b = 0,1 \;, \\
    R_{ab} &= -\left( c_1^2 + \dots + c_{8}^2 - c_9^2 - \dots - c_{16}^2 \right) \delta_{ab} \eqdef r_2\delta_{ab} \;, \qquad\quad\; a,b = 2,3 \;.
\end{align}\end{subequations}
The remaining Einstein equations, in conjunction with \eqref{betaMetric2}, imply the additional constraints
\begin{fleqn}\begin{subequations}\label{CconstraintsM4CS2CS2CS2_2_IIB}\begin{equation}
    0 = c_1 c_5 + c_2 c_6 + c_3 c_7 + c_4 c_8 + c_9 c_{13} + c_{10} c_{14} + c_{11} c_{15} + c_{12} c_{16} \;,
\end{equation}\begin{equation}
    0 = c_1 c_3 + c_2 c_4 + c_5 c_7 + c_6 c_8 + c_9 c_{11} + c_{10} c_{12} + c_{13} c_{15} + c_{14} c_{16} \;,
\end{equation}\begin{equation}
    0 = c_1 c_2 + c_3 c_4 + c_5 c_6 + c_7 c_8 + c_9 c_{10} + c_{11} c_{12} + c_{13} c_{14} + c_{15} c_{16} \;,
\end{equation}\begin{equation}\begin{split}
    \mu &= c_1^2 + c_2^2 + c_3^2 + c_4^2 - c_5^2 - c_6^2 - c_7^2 - c_8^2 + c_9^2 + c_{10}^2 + c_{11}^2 + c_{12}^2 \\
    &- c_{13}^2 - c_{14}^2 - c_{15}^2 - c_{16}^2 \;,
\end{split}\end{equation}\begin{equation}\begin{split}
    \mu &= c_1^2 + c_2^2 - c_3^2 - c_4^2 + c_5^2 + c_6^2 - c_7^2 - c_8^2 + c_9^2 + c_{10}^2 - c_{11}^2 - c_{12}^2 \\
    &+ c_{13}^2 + c_{14}^2 - c_{15}^2 - c_{16}^2 \;,
\end{split}\end{equation}\begin{equation}\begin{split}
    \mu &= c_1^2 - c_2^2 + c_3^2 - c_4^2 + c_5^2 - c_6^2 + c_7^2 - c_8^2 + c_9^2 - c_{10}^2 + c_{11}^2 - c_{12}^2 \\
    &+ c_{13}^2 - c_{14}^2 + c_{15}^2 - c_{16}^2 \;.
\end{split}\end{equation}\end{subequations}\end{fleqn}
In addition, from \eqref{curvaturesM4CS2CS2CS2_IIB} and \eqref{curvatureM4_222} we find 
\begin{equation}
    r_2 = r_1 - \frac32\nu \;.
\end{equation}
Let us now illustrate a few examples of solutions to the above system of constraints.

\subsubsection*{Example 1: $\cM_4 \simeq \rm AdS_2\times S^2$}
\label{Example3131}
A first solution can be found by setting
\begin{tcolorbox}[colback=blue!50!green!5,colframe=black,title={},width=1.01\textwidth,left=0pt, right=0pt]\begin{equation}\begin{gathered}
    c_5 = \frac12 s_5 \sqrt{\frac{3\nu(4\mu + 3\nu)}{4\mu + 6\nu}} \;,\quad
    c_8 = s_8 \frac{3\nu}{2\sqrt{4\mu + 6\nu}} \;, \\[5pt]
    c_9 = s_9 \frac{4\mu + 3\nu}{2\sqrt{4\mu + 6\nu}} \;,\quad
    c_{12} = \frac12 s_{12} \sqrt{\frac{3\nu(4\mu + 3\nu)}{4\mu + 6\nu}} \;,\quad
    r_1 = \mu + \frac{3\nu}{2} \;,\quad 
    r_2 = \mu \;,
\end{gathered}\end{equation}\end{tcolorbox}
\noindent and taking the rest of the $c_i$ parameters to be zero. Also, $s_i=\pm1$ and $s_8 s_9 = s_5 s_{12}$. For the four-dimensional space $\cM_4$ we can assume the decomposition $\rm AdS_2 \times S^2$, in agreement with \eqref{dconstraintsM4CS2CS2CS2_5}. The $\rm AdS_2$ extends along $(e^0 , e^1)$ and is normalised such that $R_{ab} = - r_1 \eta_{ab}$, while $\rm S^2$ extends in $(e^2 , e^3)$ and follows the normalisation $R_{ab} = r_2 \delta_{ab}$. The resulting ten-dimensional background has geometry of the form $\rm AdS_2 \times S^2 \times CS^2_\lambda \times CS^2_\lambda \times CS^2_\lambda$ and is real for all values of $\lambda$ in $[0 , 1)$.

\subsubsection*{Example 2: $\cM_4 \simeq \rm AdS_2\times T^2$}
\label{Example3132}
Another interesting case arises by imposing $r_2 = 0$. The Ricci tensor components \eqref{curvaturesM4CS2CS2CS2_IIB} in this case suggest that $\cM_4$ can split as $\cM^t_2 \times {\rm T^2}$. Here $\cM^t_2$ is an Einstein space of negative constant curvature extending in $(e^0 , e^1)$, and the torus $\rm T^2$ extends in $(e^2 , e^3)$. The space $\cM^t_2$ can be safely chosen to be $\rm AdS_2$, normalised such that $R_{ab} = - r_1 \eta_{ab}$. A solution for the parameters $c_i$ is
\begin{tcolorbox}[colback=blue!50!green!5,colframe=black,title={},width=1.01\textwidth,left=0pt, right=0pt]\begin{equation}\label{SolSec313Ex2}\begin{gathered}
    c_1 = s_1 \sqrt{\frac{3 \nu}{8} + \frac{\mu}{2}} \;, \qquad
    c_8 = s_8 \sqrt{\frac{3 \nu}{8} - \frac{\mu}{2}} \;, \\[5pt]
    c_{11} = s_{11} \sqrt{\frac{3 \nu}{8}} \;, \qquad
    c_{14} = s_{14} \sqrt{\frac{3 \nu}{8}} \;, \qquad
    r_1 = \frac{3\nu}2 \;.
\end{gathered}\end{equation}\end{tcolorbox}
\noindent The result is real for any $\lambda \in [0,1)$, with $s_i=\pm1$, and describes a geometry of the type $\rm AdS_2 \times T^2 \times CS^2_\lambda \times CS^2_\lambda \times CS^2_\lambda$.

\subsubsection*{Comment}
It is worth pointing out that the above solution is not the T-dual of the solution in eq.~\eqref{SolSec312}. Of course, in the present example one may solve the type-IIB equations of motion by choosing the parameters $c_i$ as in eq.~\eqref{SolSec312}; in that case, the two backgrounds become T-dual to each other. Conversely, the example in Sec.~\ref{Subsubsec:IIA_M4CS2CS2CS2} admits an alternative solution with parameters $c_i$ given in eq.~\eqref{SolSec313Ex2}, which corresponds to the T-dual of that solution.

\subsubsection*{Example 3: $\cM_4 \simeq \rm AdS_2\times H_2$}
\label{Example3133}
A third possibility is to set
\begin{tcolorbox}[colback=blue!50!green!5,colframe=black,title={},width=1.01\textwidth,left=0pt, right=0pt]\begin{equation}\begin{gathered}
    c_3 = s_3 \sqrt{\frac{3\nu}8} \;, \qquad
    c_6 = s_6 \sqrt{\frac{3\nu}8} \;, \qquad
    c_9 = s_9 \sqrt{\mu} \;, \\[5pt]
    r_1 = \frac14(4\mu + 3\nu) \;, \qquad
    r_2 = \frac14(4\mu - 3\nu) \;,
\end{gathered}\end{equation}\end{tcolorbox}
\noindent with $s_i = \pm1$ and the rest of the constants $c_i$ being zero. The solution is again real for all values of the deformation parameter $0\leq\lambda<1$. However, it's worth pointing out that $r_1 > 0$ and $r_2 < 0$ for $\lambda \in [0 , 1)$. This observation, together with eq. \eqref{dconstraintsM4CS2CS2CS2_5} and the structure of the Ricci tensor \eqref{curvaturesM4CS2CS2CS2_IIB} on $\cM_4$, allow us to choose $\cM_4 = {\rm AdS_2 \times H_2}$. Therefore, we have found a ten-dimensional background with geometry $\rm AdS_2 \times H_2 \times CS^2_\lambda \times CS^2_\lambda \times CS^2_\lambda$.

\subsubsection*{Example 4: $\cM_4 \simeq \rm AdS_4$}
\label{Example3134}
Finally, one can also aim for a solution with an $\rm AdS_4$ factor in the geometry. In view of \eqref{M4CS2CS2CS2IIBRRAnsatz}, \eqref{dconstraintsM4CS2CS2CS2_5} and \eqref{curvaturesM4CS2CS2CS2_IIB}, this can be achieved by switching off the self-dual five-form and setting $r_2 = - r_1$. A solution in this case is determined by setting
\begin{tcolorbox}[colback=blue!50!green!5,colframe=black,title={},width=1.01\textwidth,left=0pt, right=0pt]\begin{equation}\begin{gathered}
    c_1 = s_1 \sqrt{\frac{3 \nu}{8} + \frac{\mu}{2}} \;, \qquad
    c_8 = s_8 \sqrt{\frac{3 \nu}{8} - \frac{\mu}{2}} \;, \qquad
    r_1 = - r_2 = \frac{3 \nu}{4} \;,
\end{gathered}\end{equation}\end{tcolorbox}
\noindent with $s_i = \pm1$ and the rest of the constants $c_i$ being zero. The parameters $c_1$ and $c_8$ are real for all values of $\lambda \in [0,1)$. The corresponding type-IIB background has geometry $\rm AdS_4 \times CS^2_\lambda \times CS^2_\lambda \times CS^2_\lambda$.


\refstepcounter{subsubsection}\subsubsection*{\thesubsubsection\quad Type-IIA on $\cM_2 \times {\rm CS}^2_\lambda \times {\rm CS}^2_\lambda \times {\rm CS}^2_\lambda \times {\rm CS}^2_\lambda$}
\label{Subsubsec:IIA_M2CS2CS2CS2CS2}

We will now continue along the lines of sections \ref{Subsubsec:IIA_M6CS2CS2}, \ref{Subsubsec:IIA_M4CS2CS2CS2} and \ref{Subsubsec:IIB_M4CS2CS2CS2} to embed four copies of $\rm CS^2_{\lambda}$ in type-IIA supergravity. The corresponding ten-dimensional geometry is given by the direct product
\begin{equation}
    \cM_2 \rm \times CS^2_\lambda \times CS^2_\lambda \times CS^2_\lambda \times CS^2_\lambda \;.
\end{equation}
The two-dimensional manifold $\cM_2$ extends in the directions $(e^0,\,e^1)$ with line element
\begin{equation}
    ds^2_{\cM_2} = - (e^0)^2 + (e^1)^2 \;.
\end{equation}
The other eight dimensions have Euclidean signature and are spanned by the frame fields $e^2 \to \mathfrak{e}^1$, $e^3 \to \mathfrak{e}^2$; $e^4 \to \mathfrak{e}^1$, $e^5 \to \mathfrak{e}^2$; $e^6 \to \mathfrak{e}^1$, $e^7 \to \mathfrak{e}^2$; and $e^8 \to \mathfrak{e}^1$, $e^9 \to \mathfrak{e}^2$.
Each copy of $\rm CS^2_\lambda$ is assigned the pair of coordinates $(x_i,\omega_i)$, with $i = 1, \dots, 4$. Furthermore, the dilaton is taken to be the sum
\begin{equation}\label{Dilaton_2222}
    \Phi = - \frac12 \log \left(\frac{2\omega_1^2}{\omega_{1+}^2}\right) - \frac12 \log \left(\frac{2\omega_2^2}{\omega_{2+}^2}\right) - \frac12 \log \left(\frac{2\omega_3^2}{\omega_{3+}^2}\right) - \frac12 \log \left(\frac{2\omega_4^2}{\omega_{4+}^2}\right) \;,
\end{equation}
while the NS two-form is as usual assumed to vanish.

The curvature of the two-dimensional manifold $\cM_2$ can be derived from the field equation for the dilaton \eqref{dilatonEOM} and the identity \eqref{betaScalar2}, giving
\begin{equation}\label{curvatureM4_2222}
    R_{\cM_2} = -4\nu \;.
\end{equation}
Like in the previous examples, the Ricci scalar of the unknown manifold is constant and negative for all values of the deformation parameter $\lambda\in[0,1)$.

To further investigate the properties of $\cM_2$ we will adopt an ansatz for the fields of the RR sector. Since only the four-form $F_4$ has large enough rank to accommodate legs in each copy of $\rm CS^2_\lambda$, we will set $F_2 = 0$ and
\begin{equation}\begin{split}
    F_4 & = 2 e^{-\Phi} \big( c_1 e^{2468} + c_2 e^{2469} + c_3 e^{2478} + c_4 e^{2479} + c_5 e^{2568} + c_6 e^{2569} \\
    &\hspace{1.1cm}+ c_7 e^{2578} + c_8 e^{2479} + c_9 e^{3468} + c_{10} e^{3469} + c_{11} e^{3478} + c_{12} e^{3479} \\
    &\hspace{1.1cm}+ c_{13} e^{3568} + c_{14} e^{3569} + c_{15} e^{3578} + c_{16} e^{3479} \big) \;.
\end{split}\end{equation}
As usual, $c_1,\dots,c_{16}$ are constant parameters which will be determined later. This choice ensures that the Bianchi eq. \eqref{RRformsEOM} for $F_4$ is automatically satisfied, due to eq. \eqref{conditionsCS2lambda}. The corresponding equation for the RR six-form implies that
\begin{equation}\label{dconstraintsM2CS2CS2CS2CS2_4}
    de^{01} = 0 \;,
\end{equation}
provided that $F_4$ is non-trivial. Moreover, the second formula in \eqref{fieldeqnH} gives
\begin{equation}\label{CconstraintsM2CS2CS2CS2CS2_1}
    0 = c_1c_{16} - c_2c_{15} - c_3c_{14} + c_4c_{13} - c_5c_{12} + c_6c_{11} + c_7c_{10} - c_8c_9 \;.
\end{equation}

The Einstein equations \eqref{EinsteinEOMs} and \eqref{EinsteinEOMsTmn}, in conjunction with \eqref{betaMetric2}, reveal the structure of the Ricci tensor on $\cM_2$, which reads
\begin{equation}\label{curvatureM2CS2CS2CS2CS2}
    R_{ab} = -\left( c_1^2 + \dots + c_{16}^2 \right) \eta_{ab} \eqdef -r\eta_{ab} \;, \qquad a,b = 0,1 \;.
\end{equation}
In other words, $\cM_2$ is an Einstein space of constant negative curvature, such that
\begin{equation}\label{rcurvatureM2CS2CS2CS2CS2}
    r = 2\nu \;.
\end{equation}
To arrive at the above result we made use of \eqref{curvatureM4_2222}. In addition, eqs. \eqref{EinsteinEOMs} and \eqref{EinsteinEOMsTmn} provide the constraints below
\begin{fleqn}\begin{subequations}\label{CconstraintsM2CS2CS2CS2CS2_2}\begin{equation}
    0 = c_1 c_9 + c_2 c_{10} + c_3 c_{11} + c_4 c_{12} + c_5 c_{13} + c_6 c_{14} + c_7 c_{15} + c_8 c_{16} \;,
\end{equation}\begin{equation}
    0 = c_1 c_5 + c_2 c_6 + c_3 c_7 + c_4 c_8 + c_9 c_{13} + c_{10} c_{14} + c_{11} c_{15} + c_{12} c_{16} \;,
\end{equation}\begin{equation}
    0 = c_1 c_3 + c_2 c_4 + c_5 c_7 + c_6 c_8 + c_9 c_{11} + c_{10} c_{12} + c_{13} c_{15} + c_{14} c_{16} \;,
\end{equation}\begin{equation}
    0 = c_1 c_2 + c_3 c_4 + c_5 c_6 + c_7 c_8 + c_9 c_{10} + c_{11} c_{12} + c_{13} c_{14} + c_{15} c_{16} \;,
\end{equation}\begin{equation}\begin{split}
    \mu &= c_1^2 + c_2^2 + c_3^2 + c_4^2 + c_5^2 + c_6^2 + c_7^2 + c_8^2 - c_9^2 - c_{10}^2 - c_{11}^2 - c_{12}^2 \\
    &- c_{13}^2 - c_{14}^2 - c_{15}^2 - c_{16}^2 \;,
\end{split}\end{equation}\begin{equation}\begin{split}
    \mu &= c_1^2 + c_2^2 + c_3^2 + c_4^2 - c_5^2 - c_6^2 - c_7^2 - c_8^2 + c_9^2 + c_{10}^2 + c_{11}^2 + c_{12}^2 \\
    &- c_{13}^2 - c_{14}^2 - c_{15}^2 - c_{16}^2 \;,
\end{split}\end{equation}\begin{equation}\begin{split}
    \mu &= c_1^2 + c_2^2 - c_3^2 - c_4^2 + c_5^2 + c_6^2 - c_7^2 - c_8^2 + c_9^2 + c_{10}^2 - c_{11}^2 - c_{12}^2 \\
    &+ c_{13}^2 + c_{14}^2 - c_{15}^2 - c_{16}^2 \;,
\end{split}\end{equation}\begin{equation}\begin{split}
    \mu &= c_1^2 - c_2^2 + c_3^2 - c_4^2 + c_5^2 - c_6^2 + c_7^2 - c_8^2 + c_9^2 - c_{10}^2 + c_{11}^2 - c_{12}^2 \\
    &+ c_{13}^2 - c_{14}^2 + c_{15}^2 - c_{16}^2 \;.
\end{split}\end{equation}\end{subequations}\end{fleqn}

The algrebraic system \eqref{CconstraintsM2CS2CS2CS2CS2_1}, \eqref{curvatureM2CS2CS2CS2CS2} and \eqref{CconstraintsM2CS2CS2CS2CS2_2} admits a solution which reads
\begin{tcolorbox}[colback=blue!50!green!5,colframe=black,title={},width=1.01\textwidth,left=0pt, right=0pt]\begin{equation}\begin{gathered}
    c_1 = s_1 \sqrt{\frac{\nu}{2} + \frac{3 \mu}{4}} \;, \qquad
    c_4 = s_4 \sqrt{\frac{\nu}{2} - \frac{\mu}{4}} \;, \\[5pt]
    c_{14} = s_{14} \sqrt{\frac{\nu}{2} - \frac{\mu}{4}} \;, \qquad
    c_{15} = s_{15} \sqrt{\frac{\nu}{2} - \frac{\mu}{4}} \;,
\end{gathered}\end{equation}\end{tcolorbox}
\noindent with $s_i = \pm1$ and the rest of the $c_i$ are set to zero. Such a solution is real for all values of $\lambda \in [0,1)$. Choosing $\cM_2 \simeq {\rm AdS_2}$, with normalisation as in \eqref{curvatureM2CS2CS2CS2CS2} and \eqref{rcurvatureM2CS2CS2CS2CS2}, the ten-dimensional geometry takes the form $\rm AdS_2 \times CS^2_\lambda \times CS^2_\lambda \times CS^2_\lambda \times CS^2_\lambda$.

\subsection{Solutions involving only ${\rm CS}^3_\lambda$}
\label{Subsec:Only_CS3}
We now move to type-IIA $\rm AdS$ backgrounds that arise from embedding two copies of the $\rm CS^3_\lambda$ model. As in the cases of the previous section, we focus on specific examples, among the many possibilities, for the four-dimensional geometry transverse to $\rm CS^3_\lambda \times CS^3_\lambda$.


\refstepcounter{subsubsection}\subsubsection*{\thesubsubsection\quad Type-IIA on $\cM_4 \times {\rm CS}^3_\lambda \times {\rm CS}^3_\lambda$}
\label{Subsubsec:IIA_M4CS3CS3}
Starting with the $\rm CS^3_\lambda$ model introduced in Sec. \ref{Subsec:CS^3_Model}, we will construct type-IIA solutions with ten-dimensional geometry $\cM_4 \times \rm CS^3_\lambda \times CS^3_\lambda$. Since the target space on $\rm CS^3_\lambda$ has Euclidean signature, the time-like direction has to sit in the unknown four-dimensional part $\cM_4$. The corresponding line-element by means of the frame $(e^0,\dots,e^3)$ reads
\begin{equation}
    ds^2_{\cM_4} = - (e^0)^2 + (e^1)^2 + (e^2)^2 + (e^3)^2 \;.
\end{equation}
As for the remaining six dimensions, we identify them as $e^4\to\mathfrak{e}^1$, $e^5\to\mathfrak{e}^2$, $e^6\to\mathfrak{e}^3$, $e^7\to\mathfrak{e}^1$, $e^8\to\mathfrak{e}^2$ and $e^9\to\mathfrak{e}^3$, with $\frak e^1$, $\frak e^2$ and $\frak e^3$ given by \eqref{Eq:frame3}. The two $\rm CS^3_\lambda$ are distinguished by adopting the two sets of coordinates $(x_1, y_1, \omega_1)$ and $(x_2, y_2, \omega_2)$, respectively. For the other fields of the NS sector we take only the dilaton to be non-vanishing. In particular we assume
\begin{equation}
    \Phi = - \frac12 \log \left(8\frac{\cA_1\omega_1^2}{\omega_{1+}^4}\right) - \frac12 \log \left(8\frac{\cA_2\omega_2^2}{\omega_{2+}^4}\right) \;,
\end{equation}
where each term is a copy of the scalar \eqref{scalarCS3}.

Like in the examples of the previous section, the content of the NS sector can shed light to the geometric properties of $\cM_4$. Indeed, the dilaton equation \eqref{dilatonEOM} together with \eqref{betaScalar3} imply
\begin{equation}\label{curvatureM4_CS3CS3}
    R_{\cM_4} = - 6\nu \;.
\end{equation}
This suggests that $\cM_4$ has constant and negative curvature for all values of $\lambda$ in $[0,1)$.

To get more into the geometric properties of $\cM_4$, we propose an ansatz for the RR sector, inspired by \eqref{conditionsCS3lambda}. More precisely
\begin{subequations}\begin{align}
    F_2 & = 2 e^{-\Phi} \, c_1 e^4 \wedge e^7 \;, \\
    F_4 & = 2 e^{-\Phi} \big( c_2 e^{5689} + c_3 e^{0147} + c_4 e^{2347} + c_5 e^{3567} + c_6 e^{3489} \big) \;,
\end{align}\end{subequations}
with $c_1, \dots, c_6$ being constant parameters to be determined. In turn, eq. \eqref{RRformsEOM} together with the relations \eqref{conditionsCS3lambda} imply
\begin{subequations}\label{dconstraintsM4CS3CS3_4}\begin{align}
    & de^{0123} = 0 \qquad\quad\,\, \text{if at least one of the}\; c_1, c_2 \;\text{is not zero;} \label{dconstraintsM4CS3CS3_4eq1} \\
    & de^{01} = de^{23} = 0 \quad\text{if at least one of the}\; c_3, c_4 \;\text{is not zero;} \label{dconstraintsM4CS3CS3_4eq2} \\
    & de^{012} = de^3 = 0 \quad\text{if at least one of the}\; c_5, c_6 \;\text{is not zero.} \label{dconstraintsM4CS3CS3_4eq3}
\end{align}\end{subequations}
Notice that either of eq. \eqref{dconstraintsM4CS3CS3_4eq2} or \eqref{dconstraintsM4CS3CS3_4eq3} implies \eqref{dconstraintsM4CS3CS3_4eq1}. In other words, eq. \eqref{dconstraintsM4CS3CS3_4eq1} is always valid as long as the RR sector is non-trivial.

On the other hand, the equation \eqref{fieldeqnH} for $H$ implies the following two constraints for the parameters $c_1 , \dots , c_4$
\begin{subequations}\label{CconstraintsM4CS3CS3_1}\begin{align}
    0 &= c_1c_4 + c_2c_3 \;, \\
    0 &= c_1c_3 - c_2c_4 \;.
\end{align}\end{subequations}

Combining the Einstein equations \eqref{EinsteinEOMs} and \eqref{EinsteinEOMsTmn} with \eqref{betaMetric3} we obtain the non-zero components of the Ricci tensor on $\cM_4$
\begin{subequations}\label{curvatureM4CS3CS3_1}\begin{align}
    R_{ab} &= -\left( c_1^2 + c_2^2 + c_3^2 + c_4^2 + c_5^2 + c_6^2 \right) \eta_{ab} \eqdef - r_1\eta_{ab} \;, \qquad a,b = 0,1 \;, \\
    R_{33} &= - c_1^2 - c_2^2 + c_3^2 + c_4^2 - c_5^2 - c_6^2 \eqdef r_2 \;, \\
    R_{44} &= - c_1^2 - c_2^2 + c_3^2 + c_4^2 + c_5^2 + c_6^2 \eqdef r_3 \;.
\end{align}\end{subequations}
The constants $r_1$, $r_2$ and $r_3$ are related due to \eqref{curvatureM4_CS3CS3} as
\begin{equation}\label{CconstraintsM4CS3CS3r1r2r3}
    2 r_1 - r_2 - r_3 = 6 \nu \; .
\end{equation}
In addition, we find the following two algebraic equations
\begin{subequations}\label{CconstraintsM4CS3CS3_2}\begin{align}
    2\mu &= c_1^2 - c_2^2 - c_3^2 + c_4^2 - c_5^2 + c_6^2 \;, \\
    2\mu &= c_1^2 - c_2^2 - c_3^2 + c_4^2 + c_5^2 - c_6^2 \; .
\end{align}\end{subequations}

In the following, we are going to present two distinct solutions of the algebraic system \eqref{CconstraintsM4CS3CS3_1}, \eqref{curvatureM4CS3CS3_1}, \eqref{CconstraintsM4CS3CS3r1r2r3} and \eqref{CconstraintsM4CS3CS3_2}.


\subsubsection*{Example 1: $\cM_4 \simeq \rm{AdS_4, \; AdS_2\times H_2}$}
\label{Example3211}
A solution of \eqref{CconstraintsM4CS3CS3_1}, \eqref{curvatureM4CS3CS3_1}, \eqref{CconstraintsM4CS3CS3r1r2r3} and \eqref{CconstraintsM4CS3CS3_2} can be achieved by setting
\begin{tcolorbox}[colback=blue!50!green!5,colframe=black,title={},width=1.01\textwidth,left=0pt, right=0pt]\begin{equation}\begin{gathered}
    c_1 = s_1 \sqrt{\frac{3\nu}{4} + \mu} \;, \quad
    c_2 = s_2 \sqrt{\frac{3\nu}{4} - \mu}\;, \quad
    r_1 = - r_2 = - r_3 = \frac{3\nu}2 > 0 \;,
\end{gathered}\end{equation}\end{tcolorbox}
\noindent with $s_i = \pm1$ and the rest of the constants $c_i$ being zero. The parameters $c_1$ and $c_2$ are real for all values of the deformation parameter $\lambda \in [0,1)$. The form of the Ricci tensor \eqref{curvatureM4CS3CS3_1} on $\cM_4$ allows us to consider the cases $\cM_4 \simeq {\rm AdS_4}$ and $\cM_4 \simeq {\rm AdS_2 \times H_2}$, which correspond to the ten-dimensional geometries $\rm AdS_4 \times CS^3_\lambda \times CS^3_\lambda$ and $\rm AdS_2 \times H_2 \times CS^3_\lambda \times CS^3_\lambda$ respectively.


\subsubsection*{Example 2: $\cM_4 \simeq \rm AdS_3 \times S^1$}
\label{Example3212}
A second interesting example is provided by setting $r_2 = - r_1$, $r_3 = 0$ and $c_3 = c_4 = 0$. The rest of the parameters are determined as
\begin{tcolorbox}[colback=blue!50!green!5,colframe=black,title={},width=1.01\textwidth,left=0pt, right=0pt]\begin{equation}\begin{gathered}
    c_1 = s_1 \sqrt{\frac\nu2 + \mu} \;, \quad
    c_2 = s_2 \sqrt{\frac\nu2 - \mu} \;, \\[5pt]
    c_5 = s_4 \sqrt{\frac\nu2} \;, \quad
    c_6 = s_6 \sqrt{\frac\nu2} \;, \quad 
    r_1 = -r_2 = 2\nu \;,
\end{gathered}\end{equation}\end{tcolorbox}
\noindent with $s_i = \pm1$ being signs. Such solution is real for all values of $\lambda \in [0,1)$. The Ricci tensor \eqref{curvatureM4CS3CS3_1} on $\cM_4$ now suggests that we are allowed to consider $\cM_4 \simeq {\rm AdS_3 \times S^1}$, in agreement with eq. \eqref{dconstraintsM4CS3CS3_4}. This choice corresponds to the ten-dimensional geometry $\rm AdS_3 \times S^1 \times CS^3_\lambda \times CS^3_\lambda$.

\subsection{Solutions involving only ${\rm CS}^4_\lambda$}
\label{Subsec:Only_CS4}
Continuing the search of $\rm AdS$ backgrounds we will consider the embedding of two copies of the $\rm CS^4_\lambda$ model in type-II supergravity. Such a configuration is quite restrictive due to the fact that the $\rm CS^4_\lambda \times CS^4_\lambda$ geometry is eight-dimensional. We will show that one can nevertheless propose a type-IIA ansatz that admits an $\rm AdS$ solution, which in this case can only be $\rm AdS_2$.


\refstepcounter{subsubsection}\subsubsection*{\thesubsubsection\quad Type-IIA on $\cM_2 \times {\rm CS}^4_\lambda \times {\rm CS}^4_\lambda$}
\label{Subsubsec:IIA_M2CS4CS4}
Having as a starting point the deformed four-sphere $\rm CS^4_\lambda$ introduced in Sec. \ref{Subsec:CS^4_Model}, we will search for type-IIA solutions with geometry of the form $\cM_2 \times \rm CS^4_\lambda \times CS^4_\lambda$.
The line-element on $\cM_2$ will be expressed in terms of the frame $(e^0, e^1)$, with $e^0$ being the time-like direction, as
\begin{equation}
    ds^2_{\cM_2} = - (e^0)^2 + (e^1)^2 \;.
\end{equation}
The remaining eight dimensions are represented by two copies of \eqref{Eq:frame4}. In particular $e^2 \to \frak{e}^1$, $e^3 \to \frak{e}^2$, $e^4 \to \frak{e}^3$, $e^5 \to \frak{e}^4$; $e^6 \to \frak{e}^1$, $e^7 \to \frak{e}^2$, $e^8 \to \frak{e}^3$ and $e^9 \to \frak{e}^4$.
To avoid confusion, each copy of $\rm CS^4_\lambda$ is labelled by the two sets of coordinates $(x_1, y_1, z_1, \omega_1)$ and $(x_2, y_2, z_2, \omega_2)$, respectively.
The Kalb-Ramond field is as usual assumed to be trivial, while the dilaton $\Phi$ will be given by the sum of the scalars \eqref{scalarCS4} for each of the $\rm CS^4_\lambda$
\begin{equation}\label{dilaton44}
    \Phi = - \frac12 \log \left(64\frac{\cA_1\cB_1\omega_1^4}{\omega_{1+}^6} \right) - \frac12 \log \left(64\frac{\cA_2\cB_2\omega_2^4}{\omega_{2+}^6} \right) \;.
\end{equation}

More about the geometry on $\cM_2$  can be inferred from the dilaton equation \eqref{dilatonEOM}. In particular, if we combine it with \eqref{betaScalar4} we find that the curvature on $\cM_2$ is given by
\begin{equation}\label{curvatureM2CS4CS4}
    R_{\cM_2} = - 12\nu \;,
\end{equation}
which is constant and negative for all values of $\lambda \in [0 , 1)$.

The precise structure of the Ricci tensor on $\cM_2$ can be inspected from the RR sector. For this reason, we will take $F_2 = 0$, while for the four-form we will adopt the following ansatz
\begin{equation}
    F_4 = 2 e^{-\Phi} \big( c_1 e^{2468} + c_2 e^{2479} + c_3 e^{3568} + c_4 e^{3579} \big) \;,
\end{equation}
with $c_1, \dots, c_4$ being constant parameters to be determined. This guarantees that the Bianchi equation \eqref{RRformsEOM} for the RR four-form is trivially satisfied due to \eqref{conditionsCS4lambda}. The corresponding equation for the RR six-form implies
\begin{subequations}\label{dconstraintsM2CS4CS4_4}\begin{align}
    de^{01} &= 0
\end{align}\end{subequations}
as long as any of the parameters $c_1, \dots, c_4$ is not zero. The four-form sources the equation of motion \eqref{fieldeqnH} for the NS three-form, which gives us a constraint
\begin{equation}\label{CconstraintsM2CS4CS4_1}
    c_1 c_4 + c_2 c_3 = 0
\end{equation}
on the parameters $c_i$.

A careful treatment of the Einstein equations \eqref{EinsteinEOMs}, \eqref{EinsteinEOMsTmn}, in view of the property \eqref{betaMetric4}, reveals that the Ricci tensor on $\cM_2$ can be expressed in terms of the parameters $c_i$ that enter the RR four-form as
\begin{equation}\label{curvatureM2CS4CS4_1}
    R_{ab} = - (c^2_1 + c^2_2 + c^2_3 + c^2_4) \eta_{ab} \eqdef - r\eta_{ab} \;, \qquad a,b = 0,1 \;. 
\end{equation}
The constant $r$ is obtained using \eqref{curvatureM2CS4CS4}, where 
\begin{equation}\label{CconstraintsM2CS4CS4r}
    r = 6\nu > 0 \;.
\end{equation}
The above allows us to identify $\cM_2 \simeq {\rm AdS_2}$. The remaining Einstein equations yield
\begin{subequations}\label{CconstraintsM2CS4CS4_2}\begin{align}
    3\mu &= c_1^2 + c_2^2 - c_3^2 - c_4^2 \;, \\
    3\mu &= c_1^2 - c_2^2 + c_3^2 - c_4^2 \;.
\end{align}\end{subequations}

The system \eqref{CconstraintsM2CS4CS4_1}, \eqref{curvatureM2CS4CS4_1}, \eqref{CconstraintsM2CS4CS4r} and \eqref{CconstraintsM2CS4CS4_2} is solved by
\begin{tcolorbox}[colback=blue!50!green!5,colframe=black,title={},width=1.01\textwidth,left=0pt, right=0pt]\begin{equation}\begin{gathered}
    c_1 = s_1 \sqrt{\frac{3}{8\nu}}\,(\mu + 2\nu) \;, \quad
    c_2 = s_2 \sqrt{\frac{3}{8\nu}\,(4\nu^2 - \mu^2)} \;, \\[5pt]
    c_3 = s_3 \sqrt{\frac{3}{8\nu}\,(4\nu^2 - \mu^2)} \;, \quad
    c_4 = s_4 \sqrt{\frac{3}{8\nu}}\,(\mu - 2\nu) \;,
\end{gathered}\end{equation}\end{tcolorbox}
\noindent with $s_i = \pm 1$ obeying $s_2 s_3 - s_1 s_4 = 0$. The parameters $c_i$ above are real for all values of $\lambda \in [0,1)$. The corresponding ten-dimensional background has geometry given by the direct product $\rm AdS_2 \times CS^4_\lambda \times CS^4_\lambda$.

\subsection{Solutions that mix ${\rm CS}^2_\lambda$ and ${\rm CS}^4_\lambda$}
\label{Subsec:CS_different_kind}
In this last section we construct solutions of the type-II supergravities which accommodate the content of $\lambda$-models of different kind. In particular we find backgrounds that mix\footnote{The term ``mix'' refers to geometries whose internal part contains deformed spheres of different dimensions, in contrast to the examples discussed in the previous sections, where the deformed spheres have the same dimension.} the $\rm CS^2_\lambda$ and $\rm CS^4_\lambda$ models.


\refstepcounter{subsubsection}\subsubsection*{\thesubsubsection\quad Type-IIA on $\cM_4 \times \rm CS^2_\lambda \times CS^4_\lambda$}
\label{Subsubsec:IIA_M4CS2CS4}
As a first example we will consider the case where the internal space takes the form $\rm CS^2_\lambda \times CS^4_\lambda$, where $\rm CS^2_\lambda$ and $\rm CS^4_\lambda$ introduced in Sec. \ref{Subsec:CS^2_Model} and in Sec. \ref{Subsec:CS^4_Model}, respectively. Therefore, the ten-dimensional geometry is $\cM_4 \times {\rm CS^2_\lambda \times CS^4_\lambda}$, and the line element on the unknown space $\cM_4$ is expressed in terms of the frame $(e^0 , \dots , e^3)$ as
\begin{equation}\label{M4CS2CS4ds2M4}
    ds^2_{\cM_4} = - (e^0)^2 + (e^1)^2 + (e^2)^2 + (e^3)^2 \;.
\end{equation}
For the internal part of the ten-dimensional geometry we identify $e^4 \to \frak{e}^1, \; e^5 \to \frak{e}^2$, with $\frak e^1, \; \frak e^2$ defined in \eqref{Eq:frame2}, and $e^6 \to \frak{e}^1$, $e^7 \to \frak{e}^2$, $e^8 \to \frak{e}^3$ and $e^9 \to \frak{e}^4$, with $\frak{e}^1, \dots, \frak{e}^4$ defined in \eqref{Eq:frame4}. As a matter of fact, the frame components $e^4$ and $e^5$ describe the geometry of $\rm CS^2_\lambda$, which we label with the coordinates $(x_1,\omega_1)$, while $e^6 , \dots , e^9$ describe the target space of $\rm CS^4_\lambda$, for which we choose the coordinates $(x_2, y_2, z_2, \omega_2)$. For the rest of the NS fields, we express the dilaton as the sum of the scalars \eqref{scalar2} and \eqref{scalarCS4}
\begin{equation}\label{dilaton24}
    \Phi = - \frac12 \log \left(\frac{2\omega_1^2}{\omega_{1+}^2}\right) - \frac12 \log \left(64\frac{\cA_2\cB_2\omega_2^4}{\omega_{2+}^6} \right) \;,
\end{equation}
while the Kalb-Ramond form is taken to be trivial.

The Ricci scalar on $\cM_4$ can be determined from the dilaton equation \eqref{dilatonEOM} using the properties \eqref{betaScalar2} and \eqref{betaScalar4}
\begin{equation}\label{curvatureM4CS2CS4}
    R_{\cM_4} = - 7\nu \;.
\end{equation}
From the above result we see that $\cM_4$ is a space of constant and negative curvature for all values of the deformation parameter $0\leq\lambda<1$.

More properties for the space $\cM_4$ arise by providing an ansatz for the RR fields. In the case of a type-IIA theory, we will only allow for a four-form which reads
\begin{equation}\begin{split}
    F_4 &= 2 e^{-\Phi} e^2 \wedge \big( c_1 e^{468} + c_2 e^{568} + c_3 e^{479} + c_4 e^{579} \big) \\
    &+ 2 e^{-\Phi} e^3 \wedge \big( c_5 e^{468} + c_6 e^{568} + c_7 e^{479} + c_8 e^{579} \big) \;.
\end{split}\end{equation}
The parameters $c_1 , \dots , c_8$ are considered to be constants which will be determined later.

This ansatz guarantees that the Bianchi equation \eqref{RRformsEOM} for $F_4$ is satisfied provided that
\begin{subequations}\label{dconstraintsM4CS2CS4_4}\begin{align}
    de^2 &= 0 \quad\text{if at least one between}\; c_1, \dots, c_4 \;\text{is not zero;} \\
    de^3 &= 0 \quad\text{if at least one between}\; c_5, \dots, c_8 \;\text{is not zero.}
\end{align}\end{subequations}
In order to see this, one has to take into account the properties \eqref{conditionsCS2lambda} and \eqref{conditionsCS4lambda}. In the same way, the field equation for the RR six-form implies that
\begin{subequations}\label{dconstraintsM4CS2CS4_6}\begin{align}
    de^{013} &= 0 \quad\text{if at least one between}\; c_1, \dots, c_4 \;\text{is not zero;} \\
    de^{012} &= 0 \quad\text{if at least one between}\; c_5, \dots, c_8 \;\text{is not zero.}
\end{align}\end{subequations}
Moreover, the equation \eqref{fieldeqnH} for the NS three-form $H$ gives a first constraint on the parameters $c_i$
\begin{equation}\label{CconstraintsM4CS2CS4_1}
    c_1 c_8 - c_2 c_7 + c_3 c_6 - c_4 c_5 = 0 \;.
\end{equation}

On the other hand, the Einstein equations \eqref{EinsteinEOMs} and \eqref{EinsteinEOMsTmn} can provide more information about the Ricci tensor on $\cM_4$. In particular, we find that its non-vanishing components are
\begin{subequations}\label{curvatureM4CS2CS4IIA_1}\begin{align}
    R_{ab} &= - \left( c_1^2 + \dots + c_8^2 \right) \eta_{ab} \eqdef - r\eta_{ab} \;, \qquad\qquad\qquad a,b = 0,1 \;, \label{curvatureM4CS2CS4_1M2}\\
    R_{22} &= - R_{33} = c_1^2 + \dots + c_4^2 - c_5^2 - \dots - c_8^2 \;, \\
    R_{23} &= 2 \left( c_1 c_5 + c_2 c_6 + c_3 c_7 + c_4 c_8 \right) \;,
\end{align}\end{subequations}
The rest of the Einstein equations imply the additional constraints
\begin{subequations}\label{CconstraintsM4CS2CS4_2}\begin{align}
    0 &= c_1 c_2 + c_3 c_4 + c_5 c_6 + c_7 c_8 \;, \\
    \mu &= c_1^2 - c_2^2 + c_3^2 - c_4^2 + c_5^2 - c_6^2 + c_7^2 - c_8^2 \;, \\
    3\mu &= c_1^2 + c_2^2 - c_3^2 - c_4^2 + c_5^2 + c_6^2 - c_7^2 - c_8^2 \;.
\end{align}\end{subequations}

At this point we would like to make the observation that a natural choice for $\cM_4$ is a split of the form $\cM_4 \simeq {\rm AdS_2 \times T^2}$, where $\rm AdS_2$ is spanned by $(e^0, e^1)$ and is normalised as in \eqref{curvatureM4CS2CS4_1M2}, while $\rm T^2$ is spanned by $(e^2, e^3)$. This is justified by the fact that we restrict our attention to geometries of the form \eqref{GeometriesInGeneral}. In particular, assuming Einstein spaces, the Ricci tensor \eqref{curvatureM4CS2CS4IIA_1} on $\cM_4$ implies that $R_{22} = R_{33} = 0$ and $R_{23} = 0$. Consequently, the subspace spanned by $(e^2 ,\, e^3)$ is flat and may be taken to be a two-torus. This is consistent with the conditions \eqref{dconstraintsM4CS2CS4_4}, which indicate that either $e^2$ and/or $e^3$ correspond to flat directions. Moreover, eq.~\eqref{curvatureM4CS2CS4_1M2} implies that the subspace along $(e^0 ,\, e^1)$ is Einstein with negative curvature and contains the time direction. It can therefore be identified with $\rm AdS_2$. In this case, eqs. \eqref{curvatureM4CS2CS4} and \eqref{curvatureM4CS2CS4IIA_1} also imply that
\begin{equation}
    r = \frac{7 \nu}{2} \, .
\end{equation}
The above system of constraints can be solved by setting
\begin{tcolorbox}[colback=blue!50!green!5,colframe=black,title={},width=1.01\textwidth,left=0pt, right=0pt]\begin{equation}\label{SolSec341}\begin{gathered}
    c_2 = s_2 \sqrt{\frac{7\nu}{8} + \frac\mu2} \;, \quad
    c_3 = s_3 \sqrt{\frac{7\nu}{8} - \frac\mu2} \;, \\[5pt]
    c_5 = s_5 \sqrt{\frac{7\nu}{8} + \mu} \;, \quad
    c_8 = s_8 \sqrt{\frac{7\nu}{8} - \mu} \;,
\end{gathered}\end{equation}\end{tcolorbox}
\noindent with $s_i$ being signs and all the other parameters vanishing. It turns out that the constants $c_2$, $c_3$, $c_5$ and $c_8$ above are real for all values of $\lambda \in [0 , 1)$. The corresponding ten-dimensional background has geometry $\rm AdS_2 \times T^2 \times CS^2_\lambda \times CS^4_\lambda$.


\refstepcounter{subsubsection}\subsubsection*{\thesubsubsection\quad Type-IIB on $\cM_4 \times \rm CS^2_\lambda \times CS^4_\lambda$}
\label{Subsubsec:IIB_M4CS2CS4}
We now move to the type-IIB analogue of the ten-dimensional space with geometry of the form $\cM_4 \times {\rm CS^2_\lambda \times CS^4_\lambda}$. The properties of the four-dimensional space $\cM_4$ will be determined later, while its line element is again given by \eqref{M4CS2CS4ds2M4}.
The other six directions transverse to $\cM_4$, as well as the rest of the NS fields, are the same as in the example of Sec. \ref{Subsubsec:IIA_M4CS2CS4}. Since both the type-IIB background here and the type-IIA one of the previous section share the same NS sector, the dilaton equation \eqref{dilatonEOM} implies \eqref{curvatureM4CS2CS4}.

In order to proceed with the construction of the solution, we also need to specify the RR fields. For this reason we assume the ansatz
\begin{subequations}\begin{align}
    F_3 &= 2 e^{-\Phi} \big( c_1 e^{468} + c_2 e^{568} + c_3 e^{479} + c_4 e^{579} \big) \;, \\
    F_5 &= 2 e^{-\Phi} (1 + \star) e^{23} \wedge \big( c_5 e^{468} + c_6 e^{568} + c_7 e^{479} + c_8 e^{579} \big) \;,
\end{align}\end{subequations}
where the one-form $F_1$ is taken to be zero and $c_1 , \dots , c_8$ are constants.

Such a choice ensures that the Bianchi equation \eqref{RRformsEOM} for $F_3$ is trivially satisfied. However, the one for the self-dual five-form implies
\begin{equation}\label{dconstraintsM4CS2CS4_5}
    de^{01} = de^{23} = 0 \;,
\end{equation}
whenever $F_5$ is non-trivial. Similarly, the Bianchi equation \eqref{RRformsEOM} for $F_7$ -- assuming that $F_3$ is non-zero -- suggests
\begin{equation}\label{dconstraintsM4CS2CS4_7}
    de^{0123} = 0 \;,
\end{equation}
which is also guaranteed by \eqref{dconstraintsM4CS2CS4_5}.

The first two constraints for the parameters $c_i$ are obtained by the equation \eqref{fieldeqnH} for the NS three-form $H$
\begin{subequations}\label{CconstraintsM4CS2CS4_1_IIB}\begin{align}
    c_1 c_5 + c_2 c_6 + c_3 c_7 + c_4 c_8 = 0 \;, \\
    c_1 c_8 - c_2 c_7 + c_3 c_6 - c_4 c_5 = 0 \;.
\end{align}\end{subequations}

Finally, elaborating on the Einstein equations \eqref{EinsteinEOMs} and \eqref{EinsteinEOMsTmn} one can shed more light on the geometric properties of $\cM_4$. Indeed we find that the non-vanishing components of the Ricci tensor on $\cM_4$ read
\begin{subequations}\label{curvatureM4CS2CS4_1_IIB}\begin{align}
    R_{ab} &= - \left( c_1^2 + \dots + c_8^2 \right) \eta_{ab} \eqdef - r_1\eta_{ab} \;, \qquad\qquad\qquad\; a,b = 0,1 \;, \\
    R_{ab} &= \left( c_5^2 + \dots + c_8^2 - c_1^2 - \dots - c_4^2 \right)\delta_{ab} \eqdef r_2\delta_{ab} \;, \qquad a,b = 2,3 \;.
\end{align}\end{subequations}
The constants $r_1$ and $r_2$ are related due to \eqref{curvatureM4CS2CS4} as
\begin{equation}
    r_1 = r_2 + \frac{7 \nu}{2} \;.
\end{equation}
In addition, taking into account the properties \eqref{betaMetric2} and \eqref{betaMetric4}, we find the extra constraints below
\begin{subequations}\label{CconstraintsM4CS2CS4_2_IIB}\begin{align}
    0 &= c_1 c_2 + c_3 c_4 + c_5 c_6 + c_7 c_8 \;, \\
    \mu &= c_1^2 - c_2^2 + c_3^2 - c_4^2 + c_5^2 - c_6^2 + c_7^2 - c_8^2 \;, \\
    3\mu &= c_1^2 + c_2^2 - c_3^2 - c_4^2 + c_5^2 + c_6^2 - c_7^2 - c_8^2 \;.
\end{align}\end{subequations}

Below we focus on few representative solutions of the algebraic system above, obeyed by the parameters $c_1 , \dots , c_8$ and $r_1 , r_2$.

\subsubsection*{Example 1: $\cM_4 \simeq \rm AdS_2 \times S^2$}
\label{Example3421}
As a first solution we consider the case where
\begin{tcolorbox}[colback=blue!50!green!5,colframe=black,title={},width=1.01\textwidth,left=0pt, right=0pt]\begin{equation}\begin{gathered}
    c_1 = s_1 \sqrt{\frac{7\nu}{8} + \mu} \;, \quad
    c_4 = s_4 \sqrt{\frac{7\nu}{8} - \mu} \;, \quad
    c_6 = s_6 \sqrt{\frac{7\nu}{8} + \mu} \;, \quad
    c_7 = s_7 \sqrt{\frac{7\nu}{8}} \;, \\[5pt]
    r_1 = \mu + \frac{7\nu}2 \;, \quad
    r_2 = \mu \;,
\end{gathered}\end{equation}\end{tcolorbox}
\noindent with $s_i$ being signs and all the other parameters are zero. Here $c_1$, $c_4$, $c_6$ and $c_7$ are real for all the values of $\lambda$ in its fundamental domain $[0,1)$. Moreover, $r_1 , \; r_2 >0$, suggesting that $\cM_4$ can be written as a direct product of two Einstein spaces, one with constant negative curvature and another one with constant positive curvature -- according to \eqref{curvatureM4CS2CS4_1_IIB}. This allows us to choose $\cM_4 \simeq {\rm AdS_2 \times S^2}$ and the corresponding ten-dimensional geometry is $\rm AdS_2 \times S^2 \times CS^2_\lambda \times CS^4_\lambda$.

\subsubsection*{Example 2: $\cM_4 \simeq \rm AdS_2 \times T^2$}
\label{Example3422}
Aiming for a solution where $\cM_4$ splits into the direct product $\cM^t_2 \times {\rm T^2}$, with the torus exteding in $(e^2 , e^3)$, we find
\begin{tcolorbox}[colback=blue!50!green!5,colframe=black,title={},width=1.01\textwidth,left=0pt, right=0pt]\begin{equation}\label{SolSec342Ex2}\begin{gathered}
    c_1 = s_1 \sqrt{\frac{7\nu}{8} + \mu} \;, \;\;
    c_4 = s_4 \sqrt{\frac{7\nu}{8} - \mu} \;, \;\;
    c_6 = s_6 \sqrt{\frac{7\nu}{8} + \frac\mu2} \;, \;\;
    c_7 = s_7 \sqrt{\frac{7\nu}{8} - \frac\mu2} \;, \\[5pt]
    r_1 = \frac{7\nu}2 \;, \quad
    r_2 = 0 \; .
\end{gathered}\end{equation}\end{tcolorbox}
\noindent Again $s_i = \pm 1$ and the rest of the $c_i$ are zero. The parameters $c_1$, $c_4$, $c_6$ and $c_7$ with the above values are real when $\lambda \in [0 , 1)$. Since $r_1 > 0$ we can safely choose $\cM^t_2 \simeq {\rm AdS_2}$ and the ten-dimensional geometry reads $\rm AdS_2 \times T^2 \times CS^2_\lambda \times CS^4_\lambda$.

\subsubsection*{Comment}
Notice that the solution above and that of eq.~\eqref{SolSec341} are not related by T-duality. The type-IIB equations of motion in the present example clearly admit an additional solution in which the parameters $c_i$ are chosen as in eq.~\eqref{SolSec341}; in that case, the two backgrounds become T-dual to each other. Conversely, the example in Sec.~\ref{Subsubsec:IIA_M4CS2CS4} admits an alternative solution with parameters $c_i$ given in eq.~\eqref{SolSec342Ex2}, which corresponds to the T-dual of that solution.

\subsubsection*{Example 3: $\cM_4 \simeq \rm AdS_2 \times H_2$}
\label{Example3423}
A third solution is provided when
\begin{tcolorbox}[colback=blue!50!green!5,colframe=black,title={},width=1.01\textwidth,left=0pt, right=0pt]\begin{equation}\begin{gathered}
    c_1 = s_1 \sqrt{\frac{7\nu}{8} + \mu} \;, \quad
    c_4 = s_4 \sqrt{\frac{7\nu}{8} - \mu} \;, \quad
    c_6 = s_6 \sqrt{\mu} \;, \\[5pt]
    r_1 = \mu + \frac{7\nu}4 \;, \quad
    r_2 = \mu - \frac{7\nu}4 \;,
\end{gathered}\end{equation}\end{tcolorbox}
\noindent with $s_i = \pm 1$ and $c_2 = c_3 = c_5 = c_7 = c_8 = 0$. It tuns out that the above non-vanishing $c_i$ are real for $\lambda \in [0 , 1)$. However, in this range $r_1 > 0$ and $r_2 < 0$, allowing us to interpret the four-dimensional space as $\cM_4 \simeq {\rm AdS_2 \times H_2}$ -- in agreement with \eqref{curvatureM4CS2CS4_1_IIB}. Hence, the corresponding ten-dimensional geometry reads $\rm AdS_2 \times H_2 \times CS^2_\lambda \times CS^4_\lambda$.

\subsubsection*{Comment: $\cM_4 \simeq \rm AdS_4$}
\label{Example3424}
One can also search for solutions where $\cM_4 \simeq \rm AdS_4$. This can be achieved by turning off the self-dual five-form. This immediately waives the restriction \eqref{dconstraintsM4CS2CS4_5} and at the same time implies that $r_1 = - r_2$ in \eqref{curvatureM4CS2CS4_1_IIB}. Nevertheless, solving the algebraic system for $c_1$, $c_2$, $c_3$, $c_4$ and $r_1$ one obtains a RR three-form with imaginary components. However, this problem arises due to the normalisations we have chosen for the geometries $\rm CS^2_\lambda$ and $\rm CS^4_\lambda$. To circumvent this, one can rescale the directions $e^4 , \dots , e^9$ as $(e^4 , e^5) \to L_2 (e^4 , e^5)$ and $(e^6 , \dots , e^9) \to L_4 (e^4 , \dots , e^9)$, where the radii $L_2$ and $L_4$ of $\rm CS^2_\lambda$ and $\rm CS^4_\lambda$ appear to be related with each other.


\refstepcounter{subsubsection}\subsubsection*{\thesubsubsection\quad Type-IIA on $\cM_2 \times \rm CS^2_\lambda \times CS^2_\lambda \times CS^4_\lambda$}
\label{Subsubsec:IIA_M2CS2CS2CS4}
We conclude with a final example that mixes two copies of $\rm CS^2_\lambda$ and $\rm CS^4_\lambda$. The expected ten-dimensional geometry is of the type
\begin{equation}
    \cM_2 \times \rm CS^2_\lambda \times CS^2_\lambda \times CS^4_\lambda \;,
\end{equation}
where the two-dimensional space $\cM_2$ is described in terms of the frame $(e^0, e^1)$ by the line element
\begin{equation}
    ds^2_{\cM_2} = - (e^0)^2 + (e^1)^2 \;.
\end{equation}
The remaining eight directions are identified as follows: $e^2 \to \frak e^1$, $e^3 \to \frak e^2$, $e^4 \to \frak e^1$ and $e^5 \to \frak e^2$, with $\frak e^1, \frak e^2$ defined in \eqref{Eq:frame2}; $e^6 \to \frak e^1$, $e^7 \to \frak e^2$, $e^8 \to \frak e^3$ and $e^9 \to \frak e^4$, with $\frak e^1, \dots, \frak e^4$ defined in \eqref{Eq:frame4}. The two copies of $\rm CS^2_\lambda$ extend in $(e^2 , e^3)$ and $(e^4 , e^5)$ and are labelled by the two sets of coordinates $(x_1, \omega_1)$ and $(x_2, \omega_2)$, respectively. As for the $\rm CS^4_\lambda$, it is spanned by $(e^6 , \dots , e^9)$ and will be labelled by the coordinates $(x_3, y_3, z_3, \omega_3)$. For the rest of the NS fields, we assume that the two-form vanishes, while the dilaton is expressed in the usual way, as the sum of the scalars \eqref{scalar2} and \eqref{scalarCS4} characterising the models involved in the internal geometry
\begin{equation}\label{dilaton224}
    \Phi = - \frac12 \log \left(\frac{2\omega_1^2}{\omega_{1+}^2}\right) - \frac12 \log \left(\frac{2\omega_2^2}{\omega_{2+}^2}\right) - \frac12 \log \left(64\frac{\cA_3\cB_3\omega_3^4}{\omega_{3+}^6} \right) \;.
\end{equation}

The value of the Ricci scalar on $\cM_2$ is again obtained from the dilaton equation \eqref{dilatonEOM}, using the properties \eqref{betaScalar2} and \eqref{betaScalar4}
\begin{equation}\label{curvatureM4CS2CS2CS4}
    R_{\cM_2} = - 8\nu \;.
\end{equation}
As a result, $\cM_2$ has constant and negative curvature for $\lambda \in [0 , 1)$.

To complete the supergravity background, we need to support the NS sector with RR fields. In the type-IIA case we are instructed by the properties \eqref{conditionsCS2lambda} and \eqref{conditionsCS4lambda} to turn on only the four-form, for which we adopt the ansatz
\begin{equation}\begin{split}
    F_4 = 2e^{-\Phi} \big( &c_1 e^{2468} + c_2 e^{2479} + c_3 e^{2568} + c_4 e^{2579} \\
    +\; &c_5 e^{3468} + c_6 e^{3479} + c_7 e^{3568} + c_8 e^{3579} \big) \;.
\end{split}\end{equation}
Here $c_1 , \dots , c_8$ are taken to be constants and will be determined later. Such a choice ensures that the Bianchi equation \eqref{RRformsEOM} for $F_4$ is trivially satisfied, while that for $F_6$ implies
\begin{equation}\label{dconstraintsM4CS2CS2CS4}
    de^{01} = 0 \;.
\end{equation}
The last holds as long as the RR sector is non-zero.

The RR four-form also sources the equation of motion \eqref{fieldeqnH} for $H$, which implies the condition
\begin{equation}\label{curvatureM4CS2CS4_1}
    0 = c_1 c_8 + c_2 c_7 - c_3 c_6 - c_4 c_5 \;.
\end{equation}

The Einstein equations \eqref{EinsteinEOMs} -- \eqref{EinsteinEOMsTmn} allow to write the non-vanishing components of the Ricci tensor $R_{ab}$ on $\cM_2$ in terms of the parameters $c_i$'s as
\begin{equation}\label{curvatureM4CS2CS4_2}
    R_{ab} = - \left( c_1^2 + \dots + c_8^2 \right) \eta_{ab} \eqdef - r\eta_{ab} \;. \qquad\qquad a,b = 0,1 \;,
\end{equation}
This shows that $\cM_2$ is also an Einstein space and we can interpret it as $\cM_2 \simeq {\rm AdS_2}$. The constant $r$ is fixed using \eqref{curvatureM4CS2CS2CS4}
\begin{equation}\label{CconstraintsM4CS2CS2CS4_3}
    r = 4\nu > 0 \;.
\end{equation}
The rest of the Einstein equations, when combined with the properties \eqref{betaMetric2} and \eqref{betaMetric4}, provide the following set of constrains on the parameters
\begin{subequations}\label{CconstraintsM4CS2CS2CS4_4}\begin{align}
    0 &= c_1 c_5 + c_2 c_6 + c_3 c_7 + c_4 c_8 \;, \\
    0 &= c_1 c_3 + c_2 c_4 + c_5 c_7 + c_6 c_8 \;, \\
    \mu &= c_1^2 + c_2^2 + c_3^2 + c_4^2 - c_5^2 - c_6^2 - c_7^2 - c_8^2 \;, \\
    \mu &= c_1^2 + c_2^2 - c_3^2 - c_4^2 + c_5^2 + c_6^2 - c_7^2 - c_8^2 \;, \\
    3\mu &= c_1^2 - c_2^2 + c_3^2 - c_4^2 + c_5^2 - c_6^2 + c_7^2 - c_8^2 \;.
\end{align}\end{subequations}

The system \eqref{curvatureM4CS2CS4_1} -- \eqref{CconstraintsM4CS2CS2CS4_4} is solved by
\begin{tcolorbox}[colback=blue!50!green!5,colframe=black,title={},width=1.01\textwidth,left=0pt, right=0pt]\begin{equation}\begin{gathered}
    c_1 = s_1 \sqrt{\nu + \frac{5\mu}4} \;, \quad
    c_4 = s_4 \sqrt{\nu - \frac{3\mu}4} \;, \quad
    c_6 = s_6 \sqrt{\nu - \frac{3\mu}4} \;, \quad
    c_7 = s_7 \sqrt{\nu + \frac\mu4} \;,
\end{gathered}\end{equation}\end{tcolorbox}
\noindent with the $s_i$ being signs and the rest of the $c_i$ zero. The non-vanishing $c_i$ are real for any value of the deformations parameter $0 \leq \lambda < 1$. From the ten-dimensional point of view, the resulting geometry reads $\rm AdS_2 \times CS^2_\lambda \times CS^2_\lambda \times CS^4_\lambda$.

\section{Conclusions}
\label{Sec:Conclusions}

In this work, we constructed various type-II supergravity backgrounds that incorporate multiple copies of the $\lambda$-models on $\nicefrac{\mathrm{SO}(3)_k}{\mathrm{SO}(2)_k}$, $\nicefrac{\mathrm{SO}(4)_k}{\mathrm{SO}(3)_k}$, and $\nicefrac{\mathrm{SO}(5)_k}{\mathrm{SO}(4)_k}$, both with and without mixing models of different dimensionality. Our approach bypasses the difficulty of solving non-linear PDEs by choosing suitable ans\"atze for the RR fields. With these ans\"atze, the type-II supergravity equations reduce to algebraic systems involving constant parameters. Moreover, the RR-field ans\"atze fully determine the structure of the resulting ten-dimensional geometries. All the examples considered here exhibit undeformed AdS factors, and they are summarized in Table~\ref{Tab:Geometry&Sugra}.

In several cases, the algebraic systems admit multiple solutions. We have presented only representative examples, as an exhaustive analysis is beyond the scope of this paper. Requiring the solutions to be real imposes bounds on the allowed values of the deformation parameter $\lambda$, which sometimes differ from the standard range $[0,1)$. Consequently, some of our backgrounds do not admit an undeformed limit ($\lambda = 0$) and/or a non-Abelian T-dual limit ($\lambda \to 1$).

Although we have presented a wide range of embeddings with geometry of the form \eqref{GeometriesInGeneral}, a careful reader may notice that certain configurations are absent. These include, for instance, embeddings that mix the models ${\rm CS}^2_{\lambda}$ and ${\rm CS}^3_{\lambda}$, or ${\rm CS}^3_{\lambda}$ and ${\rm CS}^4_{\lambda}$. One reason for their absence is that we were unable to construct ansätze containing a sufficient number of free parameters to satisfy -- without being overconstrained by -- the conditions imposed by the supergravity equations of motion. In other cases, although a formal ansatz could be written down, the resulting solutions failed to be real. Within the assumptions adopted in this work, we believe that we have exhausted all possible combinations of ${\rm CS}^{n_1}_{\lambda} \times \ldots \times {\rm CS}^{n_k}_{\lambda}$ that lead to a consistent ansatz. Interestingly, $\rm AdS_5$ solutions are absent both in the present work and in previous studies \cite{Itsios:2019izt,Itsios:2024gqr}, where only embeddings with single copies of $\rm CS^n_\lambda$ ($n = 2, 3, 4$) and their non-compact versions were considered.

The supergravity backgrounds constructed here most likely preserve no supersymmetry. This expectation stems from the intuition that the deformation of the internal geometry breaks its isometries completely. A systematic supersymmetry analysis, however, is left for future work. It would also be interesting to investigate the stability of these deformed backgrounds, which may further constrain the admissible range of $\lambda$. The same ans\"atze can be used to embed $\lambda$-models on the non-compact cosets $\nicefrac{\mathrm{SO}(1,2)_{-k}}{\mathrm{SO}(2)_{-k}}$, $\nicefrac{\mathrm{SO}(1,3)_{-k}}{\mathrm{SO}(3)_{-k}}$, and $\nicefrac{\mathrm{SO}(1,4)_{-k}}{\mathrm{SO}(4)_{-k}}$, which can be obtained from the compact cases by sending $\omega \to i \omega$ and $k \to -k$.

This work focuses on a particular class of $\lambda$-models, namely those defined on cosets based on orthogonal groups. Such models exhibit properties like those described in eqs.~\eqref{betaFuncs2}, \eqref{conditionsCS2lambda}, \eqref{betaFuncs3}, \eqref{conditionsCS3lambda}, \eqref{betaFuncs4}, and \eqref{conditionsCS4lambda}, which facilitate the construction of suitable ans\"atze for the supergravity fields. It would nevertheless be interesting to go beyond this class and investigate other deformed cosets, which are not necessarily endowed with such simplifying features. One example is the $\lambda$-model on $\nicefrac{\rm SU(3)_k}{\rm U(2)_k}$, studied in \cite{Itsios:2025fkx}, whose embedding into type-II supergravity remains an open problem. The aforementioned deformed cosets have the advantage of a vanishing antisymmetric field, which renders their supergravity realisation comparatively simpler than that of $\lambda$-models defined on groups. Supergravity backgrounds based on $\lambda$-deformations of groups have already been constructed in the cases of $\rm SU(2)$ and $\rm SL(2,\mathbb{R})$ in \cite{Itsios:2023kma} and \cite{Itsios:2023uae}. However, all known examples involve deformed AdS factors in their geometries. It would therefore be worthwhile to explore the possibility of embeddings featuring undeformed AdS factors, perhaps by considering geometries of warped-product type.

Another compelling direction is to determine whether the backgrounds constructed here arise as near-horizon limits of brane intersections. As a first step in this direction, one could study the Page charges associated with these backgrounds, which may provide insight into the underlying brane configurations. One could then attempt to formulate ans\"atze for the corresponding brane setups that are consistent with the type-II supergravity equations of motion and reproduce the geometries constructed here. Alternatively, one may investigate whether our solutions fit into classes of type-II backgrounds already known to arise as near-horizon limits of brane intersections. This occurs, for example, in the case of the $\lambda$-deformation of $\rm AdS_3 \times S^3 \times T^4$ constructed in \cite{Itsios:2023kma}, which was shown to belong to the class of type-IIB geometries identified in \cite{Legramandi:2023fjr}. Nevertheless, this is a highly non-trivial task, and none of the aforementioned approaches are guaranteed to succeed.

In this work, we have demonstrated how a certain family of integrable non-linear $\sigma$-models can find room in type-II supergravity theories. However, integrability at the level of the full supergravity background is not guaranteed and remains to be investigated. Further motivation for pursuing this line of research comes from the observation that some of the solutions constructed here may capture non-Abelian T-duals of known integrable supergravity backgrounds in the limit $\lambda \to 1$. A prominent example is the geometry $\rm AdS_3 \times S^1 \times CS^3_{\lambda} \times CS^3_{\lambda}$ discussed in Sec.~\ref{Subsubsec:IIA_M4CS3CS3}. Indeed, in the limit $\lambda \to 1$, one obtains the non-Abelian T-dual of the type-IIA background with geometry $\rm AdS_3 \times S^1 \times S^3 \times S^3$, supported by a RR four-form flux. A type-IIB version of this background can be obtained by performing a T-duality along the $\rm S^1$ direction. The resulting background preserves $16$ supercharges and is known to be integrable~\cite{Babichenko:2009dk}. In this case, the non-Abelian T-duality is performed on both three-spheres, each viewed as the symmetric coset space $\nicefrac{\rm SO(4)}{\rm SO(3)}$. The transformation acts on the $\rm SO(4)$ isometry groups of the three-spheres, completely breaking their isometries and potentially reducing the supersymmetry of the background. From the perspective of the $\lambda$-deformed background, this limit is realised by taking
\begin{equation}
    \begin{aligned}
    & \omega_1 = \frac{v_1}{2 \sqrt{2} k} \, , \qquad
    x_1 = \frac{v_2}{2 \sqrt{2} k} \, , \qquad
    y_1 = 1 - \frac{v^2_2 + v^2_3}{16 k^2} \, , \\
    & \omega_2 = \frac{u_1}{2 \sqrt{2} k} \, , \qquad
    x_2 = \frac{u_2}{2 \sqrt{2} k} \, , \qquad
    y_2 = 1 - \frac{u^2_2 + u^2_3}{16 k^2} \, , \\
    & \lambda = 1 - \frac{1}{k} \, ,
    \end{aligned}
\end{equation}
followed by sending $k \to \infty$.

Finally, in line with the original motivation, it would be worthwhile to explore the holographic duals of the type-II geometries we have obtained. This could be achieved by analysing observables on the gravity side of the AdS/CFT correspondence and interpreting them through the holographic dictionary. Relevant observables include the entanglement entropy, Wilson and \textquoteright t Hooft loops, the central charge, Page charges, and the mass spectra of mesonic operators, among others. Such an analysis could pave the way for formulating new paradigms of holographic duality, analogous to those in the context of non-Abelian T-duality~\cite{Sfetsos:2010uq,Itsios:2013wd,Lozano:2016kum,Lozano:2016wrs,Lozano:2017ole,Itsios:2017cew,Itsios:2017nou,Itsios:2016ooc}.

\section*{Acknowledgments}              
We would like to thank Olaf Hohm for reading the manuscript and useful comments. We would also like to thank R. Bonezzi, M.F. Kallimani and C. Lavino for discussions and collaborations on closely related topics.

\noindent The work of G.C.~is funded by the DFG - Projektnummer 417533893/GRK2575 “Rethinking Quantum Field Theory”. G.I.~is supported by the Einstein Stiftung Berlin via the Einstein International Postdoctoral Fellowship program “Generalised dualities and their holographic applications to condensed matter physics” (project number IPF-2020-604).

\appendix

\section{Equations of motion of type-II supergravities}
\label{Sec:Appendix_Eoms}
In this appendix we summarise the equations of motion of type IIA and type IIB supergravities following the conventions of \cite{Itsios:2024gqr}. In doing so, we are going to adopt the so-called \emph{democratic formalism} \cite{Bergshoeff:2001pv}, where the higher RR forms are related to the lower ones through
\begin{equation}
    F_p = (-1)^{\lfloor \nicefrac{p}{2} \rfloor}\star F_{10-p} \;.
\end{equation}
Odd values of $p$ refer to type IIB supergravity, while the even ones refer to type IIA.

The Einstein equations can be written as
\begin{equation}\label{EinsteinEOMs}
    R_{\mu\nu} + 2\,\nabla_\mu\nabla_\nu\Phi - \frac14 (H^2)_{\mu\nu} = \cT_{\mu\nu}
\end{equation}
where the stress-energy tensor $\cT_{\mu\nu}$ is defined by means of the RR forms as
\begin{equation}\label{EinsteinEOMsTmn}
    \cT_{\mu\nu} \defeq \frac{e^{2\Phi}}{4} \sum_{p\leq10} \left( \frac{1}{(p-1)!}(F_p^2)_{\mu\nu} - \frac1{2\cdot p!} g_{\mu\nu}\, F_p^2 \right) \;.
\end{equation}
It is understood that for type IIB $p$ takes the values $1,3,5,7,9$, while for type IIA takes the values $2,4,6,8$.
The dilaton equation depends only on the fields contained in the NS sector and reads
\begin{equation} \label{dilatonEOM}
    R + 4\,\nabla^2\Phi - 4\,(\del\Phi)^2 - \frac{H^2}{12} = 0 \;,
\end{equation}
with $H=dB$ being the field-strength of the Kalb-Ramond field. The three-form satisfies the following Bianchi and field equations
\begin{equation}\label{fieldeqnH}
    dH=0 \;, \quad d(e^{-2\Phi }H) = \frac12\sum_{p\leq 8} F_p\wedge\star F_{p+2} \;.
\end{equation}
Finally, the RR forms obey
\begin{equation}\label{RRformsEOM}
    dF_{p+2} = H\wedge F_p \;,
\end{equation}
where again, the distinction between type IIA and type IIB corresponds to even and odd values of $p$, respectively.

\section{Single copies of $\rm CS^n_\lambda$}
\label{Sec:Appendix_Single_CS2s}
In this appendix, we construct solutions of the type-II supergravities with geometry of the form $\cM_{6 - n} \times \mathbb{CP}^2 \times {\rm CS^n_\lambda}$ (for $n = 2, 4$), which are missing from the literature. Backgrounds with geometry $\cM_3 \times \mathbb{CP}^2 \times {\rm CS}^3_\lambda$ are not included here, as we were unable to construct ans\"atze with a sufficient number of parameters to satisfy the supergravity equations of motion. Before explaining the details of the constructions, we will review the geometry of $\mathbb{CP}^2$. The complex projective space $\mathbb{CP}^2$ is a four-dimensional Einstein manifold of positive constant curvature. To describe its metric, we will make use of the following frame
\begin{equation}\label{Eq:frameCP2}\begin{aligned}
    \ell_1 &= L \, d\alpha_1 \;, \quad
      \ell_2 = \frac{L}{2} \sin\alpha_1 \, d\alpha_2 \;, \quad
      \ell_3 = \frac{L}{2} \sin\alpha_1 \sin\alpha_2 \, d\alpha_3 \;,\\[5pt]
    \ell_4 &= \frac{L}{2} \sin\alpha_1 \cos\alpha_1 \, \big( d\alpha_4 + \cos\alpha_2 \, d\alpha_3 \big) \; ,
\end{aligned}\end{equation}
where the constant $L$ plays the r\^ole of a radius. The corresponding line element is
\begin{equation}
    ds^2 = \ell^2_1 + \ell^2_2 + \ell^2_3 + \ell^2_4 \; .
\end{equation}
In these conventions, the metric on $\mathbb{CP}^2$ is normalised such that $R_{\mu\nu} = \frac{6}{L^2} \, g_{\mu\nu}$. Moreover, the $\mathbb{CP}^2$ space is equipped with a two-form $\mathcal{J}_2$ which satisfies the following properties
\begin{equation}\label{CP2_2form}
    \mathcal{J}_2 \defeq \frac{1}{\sqrt2} \big( \ell_1 \wedge \ell_4 - \ell_2 \wedge \ell_3 \big) \;, \quad
    d\mathcal{J}_2 = 0 \;, \quad
    \star\mathcal{J}_2 = -\mathcal{J}_2 \;, \quad
    \mathcal{J}_2 \wedge \mathcal{J}_2 = - \text{Vol}(\mathbb{CP}^2) \;.
\end{equation}
The above properties will prove useful for defining consistent ans\"atze for the RR fields of the supergravity solutions discussed below.

\setcounter{subsection}{1}
\refstepcounter{subsubsection}\subsubsection*{\thesubsubsection\quad Type-IIB on $\cM_4 \times {\bbC\bbP}^2 \times {\rm CS}^2_\lambda$}
\label{Subsubsec:IIB_M4CP2CS2}
In the first example we will assume a geometry of the form $\cM_4 \times {\bbC\bbP}^2 \times {\rm CS}^2_\lambda$, where the $\rm CS^2_\lambda$ space has been introduced in Sec. \ref{Subsec:CS^2_Model}. The metric of the external four-dimensional space $\cM_4$ is expressed in terms of the frame $(e^0,\dots,e^3)$ as
\begin{equation}
    ds^2_{\cM_4} = - (e^0)^2 + (e^1)^2 + (e^2)^2 + (e^3)^2 \;.
\end{equation}
The remaining six dimensions are identifies as follows: $e^4 \to \ell_1$, $e^5 \to \ell_2$, $e^6 \to \ell_3$ and $e^7 \to \ell_4$, with $\ell_1, \dots, \ell_4$ defined as in \eqref{Eq:frameCP2}; $e^8 \to \frak e^1$ and $e^9 \to \frak e^2$, with $\frak e^1, \frak e^2$ defined in \eqref{Eq:frame2}. Furthermore, we assume that the Kalb-Ramond field is trivial and that the dilaton is given by the scalar \eqref{scalar2} of the $\rm CS^2_\lambda$ model. The first piece of information about $\cM_4$ comes from the dilaton equation \eqref{dilatonEOM} combined with eq. \eqref{betaScalar2}. These imply that the Ricci scalar on $\cM_4$ is given by
\begin{equation}\label{curvatureM4CP2CS2}
    R_{\cM_4} = - \frac{24}{L^2} - \nu \;,
\end{equation}
where the term $24L^{-2}$ corresponds to the curvature of the $\bbC\bbP^2$ space in our conventions. The above result reveals that $\cM_4$ is a space of constant negative curvature for all values of $\lambda$ in $[0 , 1)$.

Aiming for a type-IIB solution, we consider the following ansatz for the RR fields
\begin{subequations}\begin{align}
    F_1 &= 2 e^{-\Phi} \big( c_1 e^8 + c_2 e^9 \big) \;, \\
    F_3 &= 2 e^{-\Phi} \Big( \cJ \wedge \big( c_3 e^8 + c_4 e^9 \big) + e^{23} \wedge \big( c_5 e^8 + c_6 e^9 \big) + e^{01} \wedge \big( c_7 e^8 + c_8 e^9 \big) \Big) \;, \\
    F_5 &= 2 e^{-\Phi} (1+\star)\; \cJ \wedge \Big( \cJ \wedge \big( c_9 e^8 + c_{10} e^9 \big) + e^{23} \wedge \big( c_{11} e^8 + c_{12} e^9 \big) \nonumber \\
    &\hspace{3.2cm}+ e^{01} \wedge \big( c_{13} e^8 + c_{14} e^9 \big) \Big) \;,
\end{align}\end{subequations}
where the parameters $c_i$, $i=1,\dots,14$, are taken to be constants. The conditions \eqref{conditionsCS2lambda} ensure that the Bianchi equation \eqref{RRformsEOM} for the one-form $F_1$ is trivially satisfied. On the other hand, the higher rank RR fields imply 
\begin{subequations}\label{dconstraintsM4CP2CS2_35}\begin{align}
    & de^{0123} = 0 \qquad\qquad\;\;\text{if at least one between}\; c_1, c_2, c_3, c_4, c_9, c_{10} \;\text{is not zero;} \\
    & de^{01} = de^{23} = 0 \qquad\,\text{if at least one between}\; c_5, \dots, c_8, c_{11}, \dots, c_{14} \;\text{is not zero} \, .
\end{align}\end{subequations}
Notice that the second line above also implies the first. Therefore, the first condition holds as long as the RR sector is non-trivial.

On the other hand, the equation of motion for the NS three-form \eqref{fieldeqnH} implies the conditions
\begin{subequations}\label{CconstraintsM4CP2CS2_1}\begin{align}
    0 &= c_1 c_5 + c_2 c_6 + c_8 c_9 - c_7 c_{10} + c_3 c_{11} + c_4 c_{12} + c_4 c_{13} - c_3 c_{14} \;, \\
    0 &= c_1 c_7 + c_2 c_8 - c_6 c_9 + c_5 c_{10} - c_4 c_{11} + c_3 c_{12} + c_3 c_{13} + c_4 c_{14} \;, \\
    0 &= c_1 c_3 + c_2 c_4 + c_3 c_9 + c_4 c_{10} + c_5 c_{11} + c_8 c_{11} + c_6 c_{12} - c_7 c_{12} + c_6 c_{13} \\
    &- c_7 c_{13} - c_5 c_{14} - c_8 c_{14} \;.
\end{align}\end{subequations}

The Einstein equations \eqref{EinsteinEOMs} and \eqref{EinsteinEOMsTmn} provide an explicit expression for the non-vanishing components of the Ricci tensor on $\cM_4$ in terms of the $c_i$'s
\begin{subequations}\label{M4CP2CS2RicciM4}\begin{align}
    R_{ab} = &- \left( c_1^2 + \dots + c_{14}^2 + 2c_{12}c_{13} - 2c_{11}c_{14} \right) \eta_{ab} \eqdef - r_1\eta_{ab} \;, & a,b = 0,1 \;, \\
    R_{ab} = &- \big( c_1^2 + c_2^2 + c_3^2 + c_4^2 - c_5^2 - c_6^2 - c_7^2 - c_8^2 + c_9^2 + c_{10}^2 & \nonumber \\
    &- c_{11}^2 - c_{12}^2 - c_{13}^2 - c_{14}^2 - 2c_{12}c_{13} + 2c_{11}c_{14} \big)\delta_{ab} \eqdef r_2\delta_{ab} \;, & a,b = 2,3 \;.
\end{align}\end{subequations}
The constants $r_1$ and $r_2$ are related through \eqref{curvatureM4CP2CS2} as
\begin{equation}\label{CconstraintsM4CP2CS2_2}
    2r_1 - 2r_2 = \frac{24}{L^2} + \nu \;.
\end{equation}
The rest of the Einstein equations, together with \eqref{betaMetric2}, imply the additional constraints\begin{subequations}\label{CconstraintsM4CP2CS2_3}\begin{align}
       0 & = c_1 c_2 + c_3 c_4 + c_5 c_6 - c_7 c_8 + c_9 c_{10} + c_{11} c_{12} + c_{11} c_{13} - c_{12} c_{14} - c_{13} c_{14} \;, \\
       \frac{6}{L^2} & = - \left( c_1^2 + c_2^2 + c_5^2 + c_6^2 - c_7^2 - c_8^2 - c_9^2 - c_{10}^2 \right) \;, \\
       \mu & = c_1^2 - c_2^2 + c_3^2 - c_4^2 + c_5^2 - c_6^2 - c_7^2 + c_8^2 + c_9^2 - c_{10}^2 + c_{11}^2 - c_{12}^2 - c_{13}^2 + c_{14}^2 \\
       &- 2c_{12} c_{13} - 2c_{11} c_{14} \;.
\end{align}\end{subequations}
In the following we are going to present some representative examples of solutions to the above system \eqref{CconstraintsM4CP2CS2_1} -- \eqref{CconstraintsM4CP2CS2_3}.

\subsubsection*{Example 1: $\cM_4 \simeq \rm AdS_2 \times S^2$}
\label{ExampleB111}
A solution is obtained by setting
\begin{tcolorbox}[colback=blue!50!green!5,colframe=black,title={},width=1.01\textwidth,left=0pt, right=0pt]\begin{equation}\begin{gathered}
    c_3 = s_3 \sqrt{\frac\mu2 + \frac\nu8} \;, \quad
    c_6 = s_6 \sqrt{\frac\mu2} \;, \quad
    c_8 = s_8 \sqrt{\mu - \frac\nu8} \;, \\[5pt]
    r_1 = 2\mu \;, \quad
    r_2 = \mu - \frac{\nu}{4} \;, \quad
    L = \frac{4\sqrt3}{\sqrt{4\mu - \nu}} \;,
\end{gathered}\end{equation}\end{tcolorbox}
\noindent with $s_i$ being signs and the rest of $c_i$ are zero. Such solution is real for $\lambda \in [ 2-\sqrt3, 1 )$. Notice that $r_1, r_2 >0$ when $\lambda \in ( 2 -\sqrt{3} , 1 )$. Therefore, eq. \eqref{M4CP2CS2RicciM4} allows us to interpret $\cM_4$ as $\rm AdS_2 \times S^2$, and the corresponding ten-dimensional geometry reads $\rm AdS_2 \times S^2 \times \mathbb{CP}^2 \times CS^2_\lambda$. When $\lambda$ takes the specific value $2 - \sqrt{3}$ we find that $r_2 = 0$ and $L$ becomes infinite. This is equivalent to saying that for this value of $\lambda$ the $\rm S^2 \times \mathbb{CP}^2$ part of the geometry becomes flat.

\subsubsection*{Example 2: $\cM_4 \simeq \rm AdS_2 \times T^2$}
\label{ExampleB112}
Another interesting example can be derived when setting $r_2 = 0$. In this case, eq. \eqref{CconstraintsM4CP2CS2_2} implies that $r_1 > 0$ for $\lambda \in [0 , 1)$. The Ricci tensor \eqref{M4CP2CS2RicciM4} then suggests that we can choose $\cM_4$ to be $\rm AdS_2 \times T^2$. Such a solution can be achieved by taking
\begin{tcolorbox}[colback=blue!50!green!5,colframe=black,title={},width=1.01\textwidth,left=0pt, right=0pt]\begin{equation}\begin{gathered}
    c_3 = s_3 \sqrt{\frac\mu2 + \frac\nu8} \;, \quad
    c_6 = s_8 \sqrt{\frac\nu8} \;, \quad
    c_8 = s_8 \sqrt{\frac\mu2} \;, \\[5pt]
    r_1 = \mu + \frac\nu4 \;, \quad
    L = \frac{4\sqrt3}{\sqrt{4\mu - \nu}} \;,
\end{gathered}\end{equation}\end{tcolorbox}
\noindent where, once again, the $s_i$ are signs and the rest of the $c_i$ are zero. This solution is real for $\lambda \in [ 2 -\sqrt{3} , 1)$ and $L$ becomes infinite when $\lambda = 2 - \sqrt{3}$. In other words, when $\lambda \in ( 2 - \sqrt{3} , 1 )$, the ten-dimensional geometry is of the form $\rm AdS_2 \times T^2 \times \mathbb{CP}^2 \times CS^2_\lambda$, while for the value $\lambda = 2 - \sqrt{3}$ one has to replace $\mathbb{CP}^2$ with a four-dimensional flat space.

\subsubsection*{Example 3: $\cM_4 \simeq \rm AdS_2 \times H_2$}
\label{ExampleB113}
We can also find examples with $r_2 < 0$. A representative choice of the parameters is
\begin{tcolorbox}[colback=blue!50!green!5,colframe=black,title={},width=1.01\textwidth,left=0pt, right=0pt]\begin{equation}\begin{gathered}
    c_3 = s_3 \sqrt{\frac\mu2 + \frac\nu8} \;, \quad
    c_8 = s_8 \sqrt{\frac\mu2 - \frac\nu8} \;, \\[5pt]
    r_1 = \mu \;, \quad
    r_2 = - \frac\nu4 \;, \quad
    L = s_L \frac{4\sqrt3}{\sqrt{4\mu - \nu}} \;,
\end{gathered}\end{equation}\end{tcolorbox}
\noindent with $s_i = \pm 1$ and the rest of the $c_i$ vanish. This solution is real for $\lambda \in [2-\sqrt3, 1)$, while $r_1 > 0$ and $r_2 < 0$. As a result, eq. \eqref{M4CP2CS2RicciM4} allows for the choice $\cM_4 \simeq {\rm AdS_2 \times H_2}$. The corresponding ten-dimensional geometry is $\rm AdS_2 \times H_2 \times \mathbb{CP}^2 \times CS^2_\lambda$, as long as $\lambda \in ( 2 - \sqrt{3} , 1 )$. In the special case where $\lambda = 2 - \sqrt{3}$, the $\mathbb{CP}^2$ radius becomes infinite and therefore $\mathbb{CP}^2$ can be replaced with a four-dimensional flat space.

\subsubsection*{Example 4: $\cM_4 \simeq \rm AdS_4$}
\label{ExampleB114}
Finally, an $\rm AdS_4$ solution can be found by taking $r_2 = - r_1$. A choice that works in this case is
\begin{tcolorbox}[colback=blue!50!green!5,colframe=black,title={},width=1.01\textwidth,left=0pt, right=0pt]\begin{equation}\begin{gathered}
    c_4 = s_4 \sqrt{\frac\nu4} \;, \quad
    c_9 = s_9 \sqrt{\mu + \frac\nu4} \;, \\[5pt]
    r_1 = - r_2 = \mu + \frac\nu2 \;, \quad
    L = \frac{2\sqrt6}{\sqrt{4\mu + \nu}} \;,
\end{gathered}\end{equation}\end{tcolorbox}
\noindent with $s_i = \pm 1$ and fixing all other parameters to zero. The constants $c_4$ and $c_9$ are real for all possible values of $\lambda$ in the fundamental domain $[0,1)$. From the ten-dimensional point of view, the resulting geometry reads $\rm AdS_4 \times \bbC\bbP^2 \times CS^2_\lambda$.


\refstepcounter{subsubsection}\subsubsection*{\thesubsubsection\quad Type-IIA on $\cM_2 \times {\bbC\bbP}^2 \times {\rm CS}^4_\lambda$}
\label{Subsubsec:IIA_M2CP2CS4}
We now move to a solution that includes a single factor of the $\rm CS^4_\lambda$ model, introduced in Sec. \ref{Subsec:CS^4_Model}. The ten-dimensional geometry that we are looking for is $\cM_2 \times {\bbC\bbP}^2 \times {\rm CS}^4_\lambda$, where the external space $\cM_2$ will be determined below. We express its line element in terms of the frame $(e^0, e^1)$ as
\begin{equation}
    ds^2_{\cM_2} = - (e^0)^2 + (e^1)^2 \;.
\end{equation}
The remaining eight dimensions are identifies as follows: $e^2 \to \ell_1$, $e^3 \to \ell_2$, $e^4 \to \ell_3$ and $e^5 \to \ell_4$, where $\ell_1, \dots, \ell_4$ have been defined in \eqref{Eq:frameCP2}; $e^6 \to \frak e^1$, $e^7 \to \frak e^2$, $e^8 \to \frak e^3$, $e^9 \to \frak e^4$, with $\frak e^1, \dots, \frak e^4$ defined as in \eqref{Eq:frame4}. We take the dilaton to be given by the scalar \eqref{scalarCS4}
and the NS two-form to be trivial. Having fixed the NS sector, we can employ eqs. \eqref{dilatonEOM} and \eqref{betaScalar2} to render the Ricci scalar on $\cM_2$ 
\begin{equation}\label{curvatureM2CP2CS4}
    R_{\cM_2} = - \frac{24}{L^2} - 6\nu
\end{equation}
Here, the term $24L^{-2}$ corresponds to the curvature of the $\bbC\bbP^2$ space. From the above result we notice that $\cM_2$ has constant and negative curvature for all values of $\lambda \in [0,1)$.

We now give an ansatz for the RR two- and four-form fields with the goal of obtaining more information about the manifold $\cM_2$
\begin{subequations}\begin{align}
    F_2 &= 2 e^{-\Phi} \big( c_1 e^{68} + c_2 e^{79} \big) \;, \\
    F_4 &= 2 e^{-\Phi} \Big( \cJ \wedge \big( c_3 e^{68} + c_4 e^{79} \big) + e^{01} \wedge \big( c_5 e^{68} + c_6 e^{79} \big) \Big) \;,
\end{align}\end{subequations}
with $c_i$ being some constant parameters to be determined. This choice ensures that the Bianchi equation for $F_2$ is satisfied in view of eq. \eqref{conditionsCS4lambda}. The Bianchi equations for the higher rank forms imply
\begin{equation}\label{dconstraintsM2CP2CS4_46}
    de^{01} = 0 \;,
\end{equation}
provided that any of the RR fields is non-zero.

The equation of motion for the NS three-form gives us the first two constraints for the parameters $c_i$
\begin{subequations}\label{CconstraintsM2CP2CS4_1}\begin{align}
    0 &= c_1 c_5 + c_2 c_6 - c_3 c_4 \;, \\
    0 &= c_1 c_3 + c_2 c_4 + c_4 c_5 + c_3 c_6 \;.
\end{align}\end{subequations}

From the Einstein equations \eqref{EinsteinEOMs} and \eqref{EinsteinEOMsTmn}, we can write down an expression for the non-vanishing components of the Ricci tensor on $\cM_2$, which read
\begin{equation}\label{M4CP2CS2RicciM2}
    R_{ab} = - \left( c_1^2 + \dots + c_{6}^2 \right) \eta_{ab} \eqdef - r\eta_{ab} \;, \qquad a,b = 0,1 \;.
\end{equation}
The above result implies that $\cM_2$ is an Einstein space of constant and negative curvature and, therefore, we can choose $\cM_2 \simeq {\rm AdS_2}$. The constant $r$ can be expressed in terms of $\nu$ and the curvature of $\mathbb{CP}^2$ using the components of the Ricci tensor and eq. \eqref{curvatureM2CP2CS4}
\begin{equation}
    r = \frac{12}{L^2} + 3\nu \;.
\end{equation}
The rest of the Einstein equations imply
\begin{equation}\label{CconstraintsM2CP2CS4_2}
    \frac{6}{L^2} = c_1^2 + c_2^2 - c_5^2 - c_6^2 \;, \qquad
    3\mu = c_1^2 - c_2^2 + c_3^2 - c_4^2 - c_5^2 + c_6^2 \;.
\end{equation}
The algebraic system \eqref{CconstraintsM2CP2CS4_1} -- \eqref{CconstraintsM2CP2CS4_2} can be solved by picking
\begin{tcolorbox}[colback=blue!50!green!5,colframe=black,title={},width=1.01\textwidth,left=0pt, right=0pt]\begin{equation}\begin{gathered}
    c_1 = \frac12 s_1 \sqrt{3\nu + 3\mu} \;, \quad
    c_4 = \frac12 s_4 \sqrt{7\nu - 5\mu} \;, \quad
    c_6 = s_6 \sqrt{\nu + \mu} \;, \\[5pt]
    r = \frac12 (\mu + 7\nu) \;, \quad
    L = \frac{2\sqrt6}{\sqrt{\mu + \nu}} \;,
\end{gathered}\end{equation}\end{tcolorbox}
\noindent where $s_i$ are signs, and leaving $c_2 = c_3 = c_5 = 0$. The proposed solution is real for all values of the deformation parameter $0 \leq \lambda < 1$ and corresponds to the ten-dimensional geometry $\rm AdS_2 \times \bbC\bbP^2 \times CS^4_\lambda$.

\providecommand{\href}[2]{#2}\begingroup\raggedright\endgroup

\end{document}